\documentclass[11pt]{article}
\usepackage{jheppub}
\usepackage[T1]{fontenc}
\usepackage{bm}
\usepackage{graphicx}
\usepackage{subfigure}
\usepackage{caption}
\usepackage[normalem]{ulem}
\usepackage{gensymb}
\usepackage{pifont}
\usepackage{bbold} 
\usepackage{comment}
\usepackage{slashed}
\usepackage{wasysym}
\usepackage{fixmath}
\usepackage{snapshot}

\definecolor{GWcolour}{rgb}{1,0,0}
\definecolor{azure}{rgb}{0.0, 0.5, 1.0}

\title{\boldmath The Electro-Weak Phase Transition at Colliders: Discovery Post-Mortem}

\author[a]{Andreas Papaefstathiou}

\author[b]{and Graham White}

\affiliation[a]{Department of Physics, Kennesaw State University, 370 Paulding Ave., Kennesaw, GA 30144, U.S.A.}
\affiliation[b]{Kavli IPMU (WPI), UTIAS, The University of Tokyo, 5-1-5 Kashiwanoha, Kashiwa, Chiba, 277-8583, Japan}

\emailAdd{apapaefs@cern.ch}
\emailAdd{graham.white@ipmu.jp}

\abstract{We explore the capabilities of a future proton collider to probe the nature of the electro-weak phase transition, following the hypothetical discovery of a new scalar particle. We focus on the real singlet scalar field extension of the Standard Model, representing the most minimal, and challenging to probe, framework that can enable a strong first-order electro-weak phase transition. By constructing detailed phenomenological methods for measuring the mass and accessible couplings of the new scalar particle, we find that a 100 TeV proton collider has the potential to explore the parameter space of the real singlet model and provide meaningful constraints on the electro-weak phase transition. We empirically find some necessary conditions for the realization of a strong first order electro-weak phase transition and conjecture that additional information, including through multi-scalar processes and gravitational wave detectors, are likely needed to gauge the nature of the cosmological electro-weak transition.  This study represents the first crucial step towards solving the inverse problem in the context of the electro-weak phase transition.} 

\begin{document} 
\maketitle
\flushbottom

\section{Introduction}\label{sec:intro}
The nature of electro-weak symmetry breaking (EWSB) is one of the central questions of next generation experiments~\cite{Ramsey-Musolf:2019lsf,Caprini:2019egz,Benedikt:2653674}. The Standard Model (SM) predicts a smooth crossover \cite{Kajantie:1996mn}, but new states at the electro-weak scale can alter this picture dramatically (see, e.g.~\cite{Pietroni:1992in,Cline:1996mga,Ham:2004nv,Funakubo:2005pu,Barger:2008jx,Chung:2010cd,Espinosa:2011ax,Chowdhury:2011ga,Gil:2012ya,Carena:2012np,No:2013wsa,Dorsch:2013wja,Curtin:2014jma,Huang:2014ifa,Profumo:2014opa,Kozaczuk:2014kva,Jiang:2015cwa,Curtin:2016urg, Vaskonen:2016yiu,Dorsch:2016nrg,Huang:2016cjm,Chala:2016ykx,Basler:2016obg,Beniwal:2017eik,Bernon:2017jgv,Kurup:2017dzf,Andersen:2017ika,Chiang:2017nmu,Dorsch:2017nza,Beniwal:2018hyi,Bruggisser:2018mrt,Ghorbani:2018yfr,Athron:2019teq,Kainulainen:2019kyp,Bian:2019kmg,Li:2019tfd,Chiang:2019oms,Xie:2020bkl,Bell:2020gug,Su:2020pjw,Ghorbani:2020xqv}, for a review see \cite{Mazumdar:2018dfl}). If cosmological electro-weak symmetry breaking occurred through a strong first-order electro-weak phase transition (SFO-EWPT), it could provide an explanation as to why there is so much more matter than anti-matter in the universe~\cite{Morrissey:2012db,White:2016nbo}. Determining the true nature of EWSB therefore requires ruling out or discovering any such new states. In the context of the minimal, but challenging to detect, real singlet extension to the SM as a test case, it has been demonstrated that a proton-proton collider with a centre-of-mass energy of 100~TeV~\cite{Kotwal:2016tex,Huang:2017jws,Chen:2017qcz,Alves:2018oct,Ramsey-Musolf:2019lsf,Alves:2019igs,Papaefstathiou:2020iag}, or perhaps even 27~TeV~\cite{Papaefstathiou:2020iag}, can exclude or discover the presence of a singlet field, provided it couples to the SM sufficiently strongly to induce a SFO-EWPT. This statement has been shown to remain true even when taking a liberal approach to the theoretical uncertainties involved~\cite{Papaefstathiou:2020iag}. A potentially complementary approach towards discovery \cite{Alves:2018oct,Alves:2018jsw, Alves:2019igs,Alves:2020bpi,Zhou:2020idp} is to observe a gravitational wave background which would be present if the transition were strong enough. Such a transition is expected to leave a signal that would be observable in the frequency range of next generation experiments including Lisa \cite{Audley:2017drz}, Decigo \cite{Kawamura:2020pcg}, atom interferometers \cite{Graham:2017pmn,Bertoldi:2019tck,Badurina:2019hst}, the Einstein telescope \cite{Maggiore:2019uih} and the Cosmic Explorer \cite{Reitze:2019iox}. \par 

Thus far, phenomenological analyses have been limited to addressing the problem of whether the parameters that predict a SFO-EWPT leave an observable signal at future colliders in the form of a heavy scalar particle~\cite{Kotwal:2016tex,Huang:2017jws,Chen:2017qcz,Alves:2018oct,Ramsey-Musolf:2019lsf,Alves:2019igs,Papaefstathiou:2020iag}. This approach would only allow for discovery of the particle in question if it exists, and would not directly address the question of its origin. Here we take a step further to explore whether such a collider can \textit{distinguish between a singlet that produces a strong transition and one that does not}, given measurements of observables following the hypothetical discovery of a heavy scalar particle. 

At a 100 TeV proton collider with 30~ab$^{-1}$ of integrated luminosity, three observables can be reconstructed over the parameter space, given a high enough statistical significance in processes where the heavy scalar is ``singly'' produced and decays into vector bosons or pairs of Higgs bosons, i.e.\ $pp \rightarrow h_2 \rightarrow VV$ ($V=Z,W$) or $pp \rightarrow h_2 \rightarrow h_1 h_1$, respectively. In particular, these processes can provide sufficient information to fix the mixing angle, the mass and one of the tri-scalar couplings, as we will demonstrate here in a ``discovery post-mortem'' exercise. We will use these measurements to develop a necessary, but not sufficient, condition on these three parameters for producing a SFO-EWPT - the viable region lives within a volume of this three-dimensional parameter space. However, we will show that there are some remaining parts of the parameter space that satisfy the condition but do not predict a SFO-EWPT, i.e.\ they would yield a smooth electro-weak transition. Therefore, to truly understand the nature of electro-weak symmetry breaking, we will argue that input from other experiments is necessary, including searches for primordial gravitational wave backgrounds. 

The article is organised as follows: in section~\ref{sec:sfoewpt} we outline the main features of the real-singlet extended SM, discuss the parameter-space categorisation that we employ and summarise the main conclusions of~\cite{Papaefstathiou:2020iag} that form the basis of the present study. In section~\ref{sec:reco} we develop the methods for reconstructing the mass of a new heavy scalar particle, the mixing angle and the triple-scalar coupling, $\lambda_{112}$. In section~\ref{sec:exploration} we apply these techniques on the SFO-EWPT parameter space to obtain the expected constraints for a selection of benchmark points. We present our conclusions and discussion in section~\ref{sec:conclusions}. Appendix~\ref{app:analyses} contains details of the phenomenological Monte Carlo analyses employed in our study. Appendix~\ref{app:xsecuncert} contains an explanation of the gaussian approximation formula that we employ to calculate the uncertainty on the signal cross section measurements.

\section{SFO-EWPT catalyzed by a real singlet scalar field}\label{sec:sfoewpt}

\subsection{Standard Model augmented by a real singlet scalar field}\label{sec:model}

When the SM is extended by a real singlet scalar field, the most general form of the scalar potential that depends on the Higgs doublet field, $H$, and a gauge-singlet scalar field, $S$, is given by (see, e.g.~\cite{OConnell:2006rsp, Profumo:2007wc, Barger:2007im, Espinosa:2011ax, Pruna:2013bma, Chen:2014ask, Kotwal:2016tex, Robens:2016xkb, Englert:2020gcp, Adhikari:2020vqo}):

\begin{eqnarray}\label{eq:xsm}
V(H,S) &=& \mu^2 (H^\dagger H) + \frac{1}{2} \lambda (H^\dagger H)^2 + K_1 (H^\dagger H) S \\ \nonumber
             &+& \frac{K_2}{2} (H^\dagger H) S^2 + \frac{M_S^2}{2} S^2 + \frac{\kappa}{3} S^3 + \frac{\lambda_S}{2} S^4 \;,
\end{eqnarray}
where the interactions proportional to $K_{1,2}$ constitute the Higgs ``portal'' that links the SM with the singlet scalar. Note that here we do not impose a $\mathbb{Z}_2$ symmetry that would preclude terms of odd 
powers of $S$. Such terms are often key in catalysing a tree-level barrier between the electro-weak symmetric and broken phases, thus resulting in a stronger transition. Indeed, the parameter-space exploration of ref.~\cite{Papaefstathiou:2020iag} demonstrates that a large fraction of the SFO-EWPT-viable points possess large scale hierarchy for $\mathbb{Z}_2$-symmetry breaking terms odd in $S$, i.e.\ $K_1/|\mu|>>1$ and $\kappa /|\mu |>>1$.

After EWSB occurs, the Higgs doublet and the singlet scalar fields both obtain vacuum expectation values (vevs) $v_0$ and $x_0$, respectively. To obtain the physical states, we expand about these: $H \rightarrow (v_0 + h) / \sqrt{2}$, with $v_0 \simeq 246$~GeV and $S \rightarrow x_0 + s$. Inevitably, the two states $h$ and $s$ mix through both the Higgs portal parameters $K_1$ and $K_2$ as well as the singlet vev and hence they do not represent mass eigenstates. Therefore, upon diagonalising the mass matrix one obtains two eigenstates,
\begin{eqnarray}
h_1 &=& h \cos \theta + s \sin \theta \;, \\ \nonumber 
h_2 &=& - h \sin \theta + s \cos \theta \;. 
\end{eqnarray}
where $\theta$ is a mixing angle that can be expressed in terms of the parameters of the model. For $\theta \sim 0$, $h_1 \sim h$ and $h_2 \sim s$. We identify the eigenstate $h_1$ with the SM-like Higgs boson state observed at the LHC, and hence set $m_1 = 125.1$~GeV. We only consider $m_2 > m_1$ here.\footnote{The case $m_1 > 2 m_2$ in the context of SFO-EWPT in the real singlet extension of the SM was investigated in ref.~\cite{Kozaczuk:2019pet}}

Note that, following the minimisation conditions for EWSB to occur and the requirement for one of the scalar particles to yield the observed Higgs boson mass, the seven coupling parameters of eq.~\ref{eq:xsm} are reduced down to five free parameters. Therefore, a complete reconstruction of the model would require, in principle, the measurement of five uncorrelated observable quantities. 

All the couplings of $h_{1,2}$ to the rest of the SM states are simply obtained by rescaling:
\begin{equation}\label{eq:rescaledcouplings}
g_{h_1 XX} = g_{hXX}^{\mathrm{SM}} \cos \theta\;,\;\; g_{h_2 XX} =  - g_{hXX}^{\mathrm{SM}}  \sin \theta\;,
\end{equation}
with $XX$ being any SM final state, i.e.\ fermions or gauge bosons. These allow for constraints to be imposed on the mixing angle $\theta$ through the measurements of SM-like Higgs boson (i.e.\ $h_1$) signal strengths and for searches of $h_2$ decaying to SM particles. In addition, if $m_2 \geq 2 m_1$, then $h_2 \rightarrow h_1 h_1$ becomes kinematically viable through the triple coupling $h_2-h_1-h_1$, given at tree level in terms of the parameters of the model by:
\begin{eqnarray}
      \lambda_{112} &=& v_0 ( K_2 - 6 \lambda )  c_\theta^2 s_\theta - \frac{1}{2} K_2 v_0 s_\theta^3 \\ \nonumber
 &+& ( -K_1 - K_2 x_0 + \kappa + 6 \lambda_S x_0 ) c_\theta s_\theta^2 + \frac{1}{2}  (K_1 + K_2 x_0 ) c_\theta^3 \;, 
\end{eqnarray}
where $V(h_1-h_1-h_2) \supset \lambda_{112} h_2 h_1 h_1$.\footnote{We note here that $\lambda_{112}$ is the actual factor that appears in the potential, i.e.\ there are no factors of 1/2 as it is sometimes conventional to include.}  In the studies of the present article, we will assume that indeed $m_2 > 2 m_1$, such that $h_2 \rightarrow h_1 h_1$ is open, with both SM-like Higgs boson $h_1$ scalars being on shell.

\subsection{Calculation of the phase transition}\label{sec:phasetrans}

Describing the nature of the electro-weak transition is an ongoing theoretical challenge. The current state-of-the-art technique is a gauge-invariant calculation at next-to-leading order (NLO) in dimensional reduction \cite{Croon:2020cgk,Gould:2021oba,Niemi:2021qvp}. To derive the dynamics of the dimensionally-reduced potential at NLO requires the calculation of ${\cal O}(10^2)$ diagrams which makes the application of the state of the art to large parameter scans in multiple models a work-in-progress. Even at NLO, for sufficiently large couplings, perturbation theory begins to struggle to make sharp predictions \cite{Gould:2021oba} and, for weak transitions, infrared divergences in the physical Higgs mode can cause perturbation theory to qualitatively disagree with lattice results \cite{Niemi:2020hto}.

In the meantime, there is significant utility in approximate methods that can estimate the nature of the electro-weak phase transition in a large multi-parameter scan, so long as one is upfront about the theoretical uncertainties in such an approach. In doing so, one must make a somewhat unfortunate choice between gauge-dependent methods, or a gauge-independent method that does not include a resummation of divergent infrared modes at leading order \cite{Patel:2011th}. The enormous unphysical scale dependence found in multiple studies due to the poor convergence of perturbation theory to ${\cal O}(g^4)$ \cite{Gould:2021oba} suggests that it is a heavy cost to neglect resummation terms at ${\cal O}(g^3)$. For a scalar singlet, unlike the standard model, the new contributions, either to a tree-level barrier or the thermal barrier, are gauge independent due to the gauge-singlet nature of the new field. We therefore follow ref.~\cite{Papaefstathiou:2020iag} in using a gauge-dependent method with leading-order Arnold-Espinosa resummation \cite{Arnold:1992rz,Arnold:1992fb}. Specifically, we include the one-loop corrections at finite temperature, evaluated in the covariant gauge using the $\overline{\mathrm{MS}}$ scheme, and include a leading-order resummation of Daisy diagrams:
\begin{equation}
    V(h,s , T, \mu , \xi_W,\xi_Z)= V_{\rm Tree}(h,s , \mu) + V_{\rm CW} (h,s, \mu , \xi_W, \xi _Z) + V_T(h,s,T, \mu , \xi_W, \xi _Z)
\end{equation}
where $\xi_{W,Z}$ are the gauge parameters, $\mu$ is the renormalization scale, $V_{\rm CW}$ is the zero-temperature one-loop Coleman-Weinberg correction and $V_T$ is the thermal potential. For details see ref.~\cite{Papaefstathiou:2020iag}.
Across the parameter space, it was found in~\cite{Papaefstathiou:2020iag} that the majority of points were only predicting a strong first-order electro-weak transition for some values of the unphysical renormalization scale. The points were categorised in terms of how robust the claim that the point predicts a strong first-order transition is:
\begin{itemize}
    \item ``Ultra-conservative'': The transition is strongly first order and the parameters reproduce zero-temperature observables for the \textit{entire range} of the renormalization scale and gauge parameters.
    \item ``Conservative'': The transition is strongly first order, independently of the gauge parameters and the renormalization scale, and we reproduce zero-temperature observables for \textit{some} values of the scale.
    \item ``Centrist'': There exists \textit{a} value of the renormalization scale and the gauge parameter where a strong first order transition is predicted and zero-temperature observables are predicted.
    \item ``Liberal'': There exists a value of the renormalization scale and gauge parameters where a strong first order transition is predicted, and a \textit{different} value where zero-temperature observables are reproduced.
\end{itemize}

\subsection{Production of a heavy SFO-EWPT scalar at colliders}\label{sec:singlefit}


In ref.~\cite{Papaefstathiou:2020iag} the capability of a 100~TeV proton collider to discover any viable SFO-EWPT parameter-space point was investigated.\footnote{There are of course multiple previous analyses of a 100 TeV proton collider probing the nature of EWSB, see refs. \cite{Kotwal:2016tex,Huang:2017jws,Chen:2017qcz,Alves:2018oct,Ramsey-Musolf:2019lsf,Alves:2019igs}. These use different, complementary, methods to the one we use here and do not include the powerful gauge-boson channels. Therefore, we focus in the rest of the paper on extending the results of ref.~\cite{Papaefstathiou:2020iag}.} Viable points were considered to be those that fall in one of the four categories defined in the previous section and that satisfied constraints coming from heavy Higgs boson searches and Higgs boson measurements, imposed via the \texttt{HiggsBounds}~\cite{Bechtle:2008jh,Bechtle:2011sb,Bechtle:2013gu,Bechtle:2015pma} and \texttt{HiggsSignals}~\cite{Bechtle:2013xfa, Stal:2013hwa, Bechtle:2014ewa}, with additional constraints coming from resonant Higgs boson pair production not included in \texttt{HiggsBounds} and the latest 13 TeV ATLAS and CMS SM-like Higgs boson global signal strengths, $\mu = \sigma^{\mathrm{measured}}/\sigma^{\mathrm{SM}}$, that were not included in \texttt{HiggsSignals}. Further details on the current collider constraints can be found in Appendix~C of~\cite{Papaefstathiou:2020iag}.

Furthermore, following detailed Monte Carlo-level phenomenological analyses for $h_2$ resonant searches, the expected statistical significance at a 100 TeV proton-proton collider, with a lifetime integrated luminosity of 30~ab$^{-1}$, was derived for the parameter-space points that appear in the four categories representing varying degrees of theoretical uncertainty,  outlined in sub-section~\ref{sec:phasetrans}. The conclusion of those studies was that a 100~TeV proton collider can efficiently discover a heavy scalar $h_2$ related to SFO-EWPT, possibly quite early in the lifetime of the experiment, through $pp \rightarrow VV$ or $pp \rightarrow h_1 h_1$ final states. This was shown to be robust against theoretical uncertainties pertaining to whether a SFO-EWPT occurs, characterised by the four categories.

The conclusions of ref.~\cite{Papaefstathiou:2020iag} thus strongly motivate the investigation of the potential measurements of the mass of the $h_2$ and the couplings involved in the $pp \rightarrow VV$ or $pp \rightarrow h_1 h_1$ processes, namely the triple coupling $\lambda_{112}$ and the mixing angle, $\theta$. We develop the necessary methods for doing so in what follows.

\section{\boldmath Reconstructing the mass, the mixing angle and $\lambda _{112}$}\label{sec:reco}

For the entire parameter space that admits a strong first-order electro-weak phase transition within theoretical uncertainties, the $pp \rightarrow h_2 \rightarrow h_1 h_1$ and $pp \rightarrow h_2 \rightarrow VV$ can be significant enough to shed light on the underlying parameters. From these observables, we can reconstruct the mixing angle $\theta$, the mass of the new scalar, $m_2$, and the effective  $h_2-h_1-h_1$ triple coupling. These measurements alone are capable of severely constraining the parameter space. In this section we outline the strategy for achieving this.

\subsection{Measuring the mass of a heavy SFO-EWPT scalar}\label{sec:massfit}

 In the case of discovery of a new scalar resonance, a global fit of the resulting signal distributions will eventually provide the ultimate measurement of its mass, $m_2$. Here we examine a subset of the final states that should provide the dominant sources of measurement of $m_2$.

The `cleanest' channel for reconstructing the mass would be $pp \rightarrow ZZ \rightarrow (\ell^+ \ell^-) (\ell'^+ \ell'^-)$, where the four final-state lepton invariant mass would provide a high-resolution measurement of $m_2$. Another viable final state is that of $pp \rightarrow h_1 h_1 \rightarrow (b\bar{b}) (\gamma \gamma)$, where all the objects are identifiable, but due to the fact that $b$-jets are involved, the resolution is expected to be somewhat worse than that of the four-lepton final state. In addition, we consider here the transverse mass observable in the $pp \rightarrow ZZ \rightarrow (\ell^+ \ell^-) (\nu \nu)$ final state, which we employ as a `fail-safe' in the cases where the other two processes fail to provide a high enough significance for mass measurement.\footnote{In contrast to the case of the SM Higgs boson, as described e.g. in ref.~\cite{Aaboud:2018wps}, the di-photon final state is likely going to be too rare to provide a sufficient number of events for the reconstruction of $m_2$.} 

We follow the analyses of $pp \rightarrow ZZ \rightarrow (\ell^+ \ell^-) (\ell'^+ \ell'^-)$, $pp \rightarrow ZZ \rightarrow (\ell^+ \ell^-) (\nu \nu)$ and $pp \rightarrow h_1 h_1 \rightarrow (b\bar{b}) (\gamma \gamma)$ as described in detail in Appendices F.1.2 and F.1.4 in ref.~\cite{Papaefstathiou:2020iag}. We outline the main features of these analyses in Appendix~\ref{app:analyses} for completeness. The details of the event generation, detector simulation and signal and background separation are identical to those described in Appendix F.1.1 of ref.~\cite{Papaefstathiou:2020iag}. We note that in the analyses of~\cite{Papaefstathiou:2020iag}, the momenta of all the
final state reconstructed objects were smeared according to
$\Delta p_T = 1.0 \times \sqrt{p_T}$ for jets ($p_T$ in GeV), $\Delta p_T = 0.2 \times \sqrt{p_T} + 0.017 \times p_T$ for photons~\cite{Kotwal:2016tex}, with $p_T$ in GeV. Muons and electron momenta are smeared according to~\cite{TheATLAScollaboration:performance1}. For jets, this implies a 10\% uncertainty at $p_T \sim 100$~GeV and 4\% for $p_T \sim 500$~GeV, which will propagate through to any mass measurements involving $b$-jets that we discuss here. Such resolutions are compatible with the best capabilities of LHC experiments, see e.g.~\cite{ATLAS:2012cse}, but will need to be re-assessed more realistically once the design of the future detectors becomes available.

In order for the mass fitting procedure to apply, we require a signal significance of at least $\Sigma = 3$ standard deviations to be achievable for any given parameter-space point, for a specific final state. The fits are performed at first instance using the four-lepton and Higgs boson pair final states, with the $pp \rightarrow ZZ \rightarrow (\ell^+ \ell^-) (\nu \nu)$ final state only employed when those fits fail due to the significance being below the chosen $\Sigma = 3$ threshold. Since the significance in $pp \rightarrow ZZ \rightarrow (\ell^+ \ell^-) (\nu \nu)$ was always found in \cite{Papaefstathiou:2020iag} to be $\Sigma \gg 5$ standard deviations for all the viable parameter space, a mass measurement can thus always be obtained for any viable parameter space point. 

\subsubsection{Mass measurement through final-state invariant masses}
To obtain the measurement of the mass $m_2$, we construct the invariant mass of the four leptons, $m(\ell_i \ell_j\ell_k \ell_l) \equiv m_{4\ell}$ in the $pp \rightarrow ZZ \rightarrow (\ell^+ \ell^-) (\ell'^+ \ell'^-)$ analysis and the invariant mass of the reconstructed $(b\bar{b})(\gamma\gamma)$ system, $m_{(b\bar{b})(\gamma\gamma)}$, in the $pp \rightarrow h_1 h_1 \rightarrow (b\bar{b}) (\gamma \gamma)$ analysis. We then perform a fit after subtracting the expected background distributions. Two fitting functions were used to model the $h_2$ signal, using the available models in the Python \texttt{lmfit} package~\cite{newville_matthew_2014_11813}: (i) either a Gaussian distribution with a peak at $m_2$ or (ii) a Skewed Voigt distribution, which is effectively the convolution of the relativistic Breit–Wigner distribution and a (skewed) Gaussian distribution peaked at $m_2$.

We assign a symmetrized estimate of the Poisson error in each bin that enters the fit, while properly taking into account the effect of the background subtraction. At each mass, we chose the distribution of the two that gave the lowest value of the reduced $\chi^2$ as calculated by the \texttt{lmfit} package. As the uncertainty of the fitted mass, we have assigned the value of the fitted $\sigma$ parameter, which corresponds to the width of the Gaussian and approximately to the width for the Skewed Voigt distribution, see the \texttt{lmfit} manual for further details. The resulting uncertainties are compatible with the error propagation of the resolutions of the object momenta involved in the mass reconstruction. We note here that a full treatment of systematic uncertainties, related e.g.\ to jet energy resolution, will be necessary in future experimental studies. 

Since the phenomenological analysis was performed for a set of pre-defined masses of the $h_2$ in $[200, 1000]$~GeV,\footnote{The minimum for the $pp \rightarrow h_1 h_1 \rightarrow (b\bar{b}) (\gamma \gamma)$ analysis was 250~GeV.} in order to obtain the expected measurement for a specific parameter-space point with an arbitrary intermediate mass and cross section, we first determine between which two masses, $m_{\mathrm{lo}}$ and $m_{\mathrm{hi}}$ the true $m_2$ lies, such that $m_{\mathrm{lo}} < m_2 < m_{\mathrm{hi}}$. We then obtain the fits at the two masses $m_{\mathrm{lo}}$ and $m_{\mathrm{hi}}$ assuming the same mixing angle $\theta$ as the parameter space point in question. The fit for $m_2$ is then approximated by:
\begin{equation}
    m_2^{\mathrm{fit}} = \omega \times (m_{\mathrm{hi}} - m_{\mathrm{lo}}) + m_{\mathrm{lo}}\;,
\end{equation}
where $\omega = (m_2 - m_{\mathrm{lo}})/(m_{\mathrm{hi}} - m_{\mathrm{lo}})$ determines the linear distance of $m_2$ from $m_{\mathrm{hi}}$ and $m_{\mathrm{lo}}$. 

The error for a given parameter-space point is estimated from these using:
\begin{equation}
    \Delta m_2^{\mathrm{fit}} = \omega \times \Delta m_{\mathrm{hi}} + (1-\omega) \times \Delta M_{\mathrm{lo}}\;,
\end{equation}
where $\Delta m_{\mathrm{hi}}$ and $\Delta M_{\mathrm{lo}}$ are the statistical errors obtained when fitting at $m_{\mathrm{hi}}$ and $m_{\mathrm{lo}}$, respectively.

\begin{figure}[htp]
  \centering
  \includegraphics[width=0.45\columnwidth]{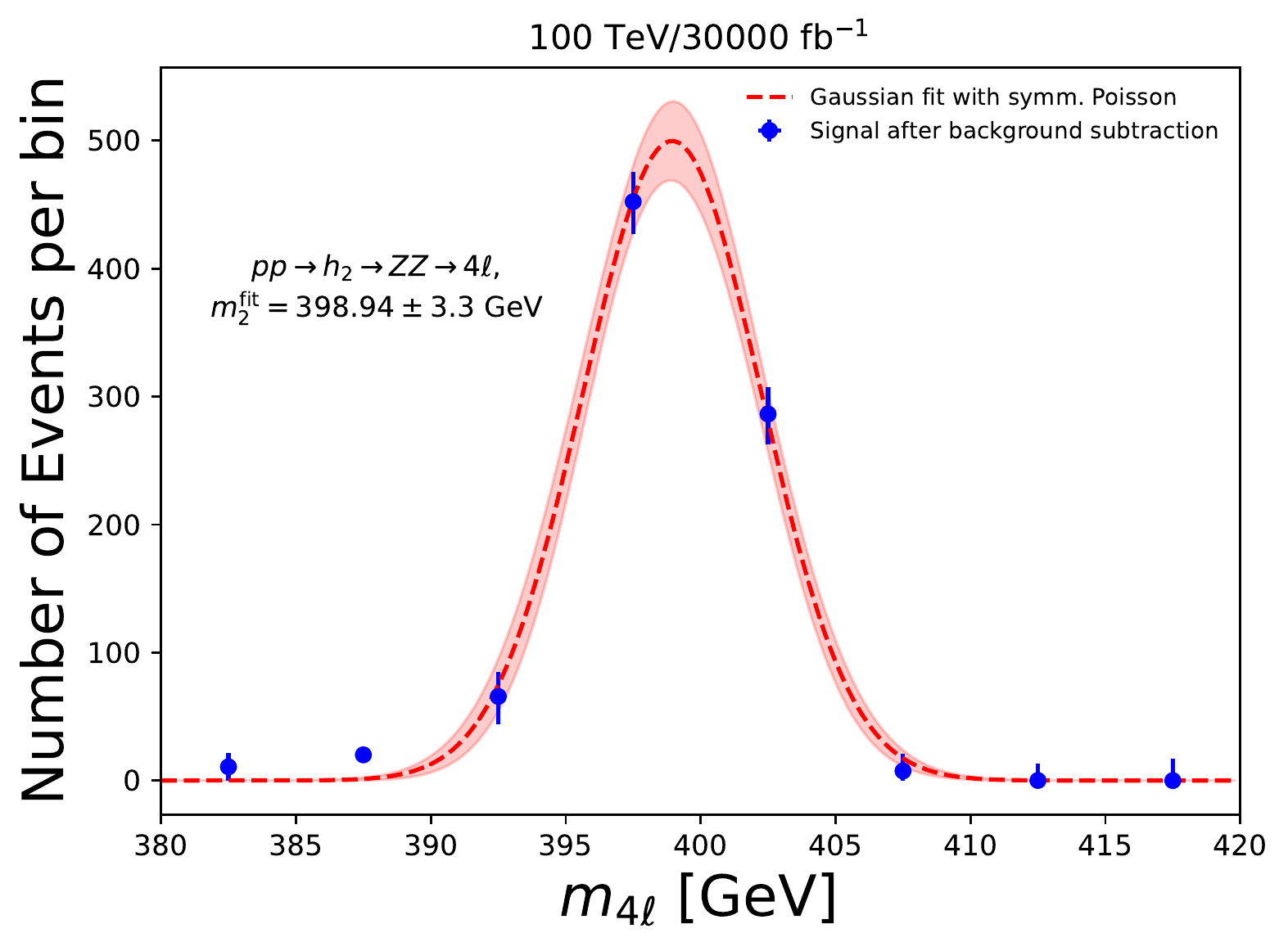}
  \includegraphics[width=0.45\columnwidth]{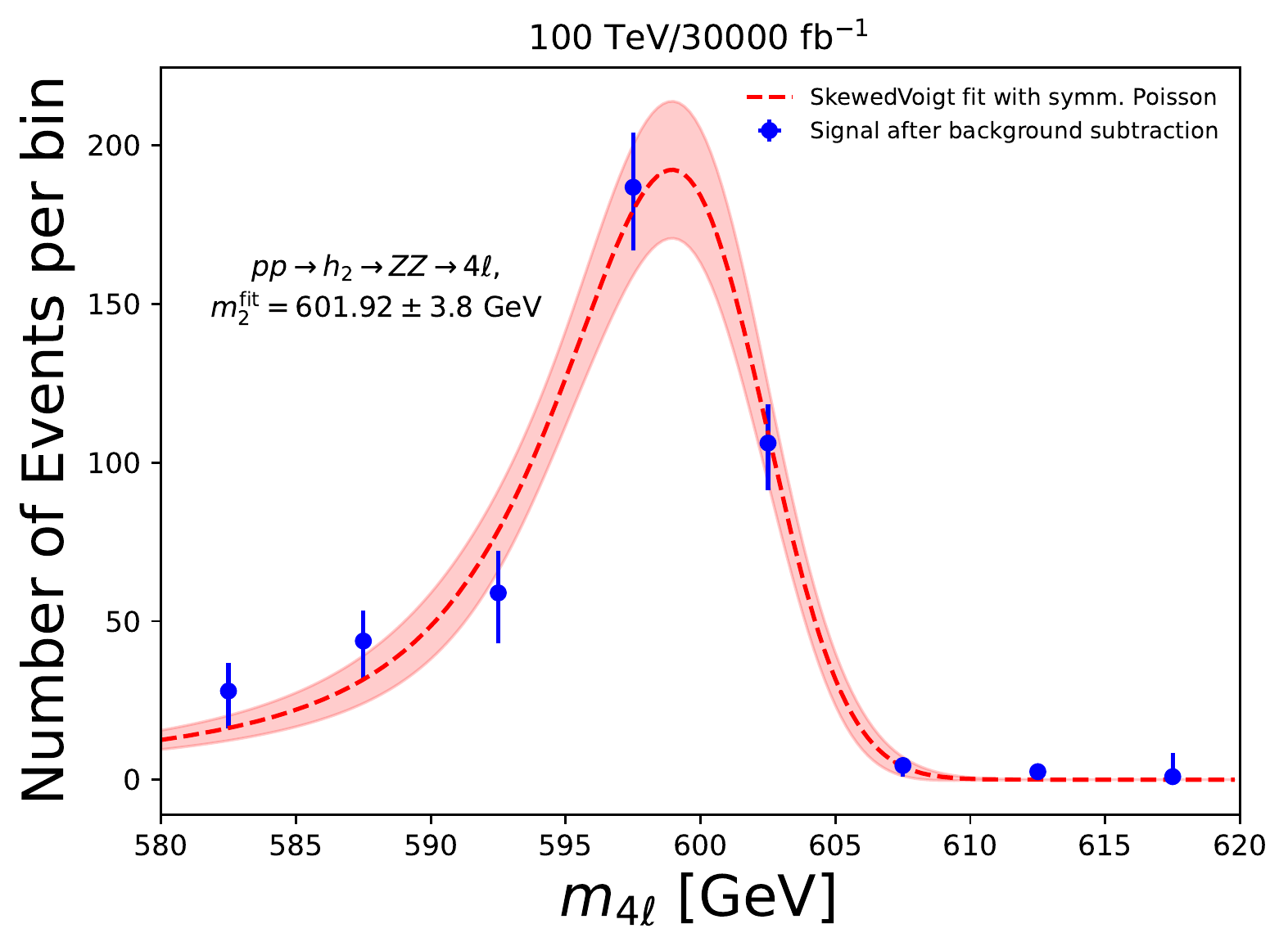}
  \includegraphics[width=0.45\columnwidth]{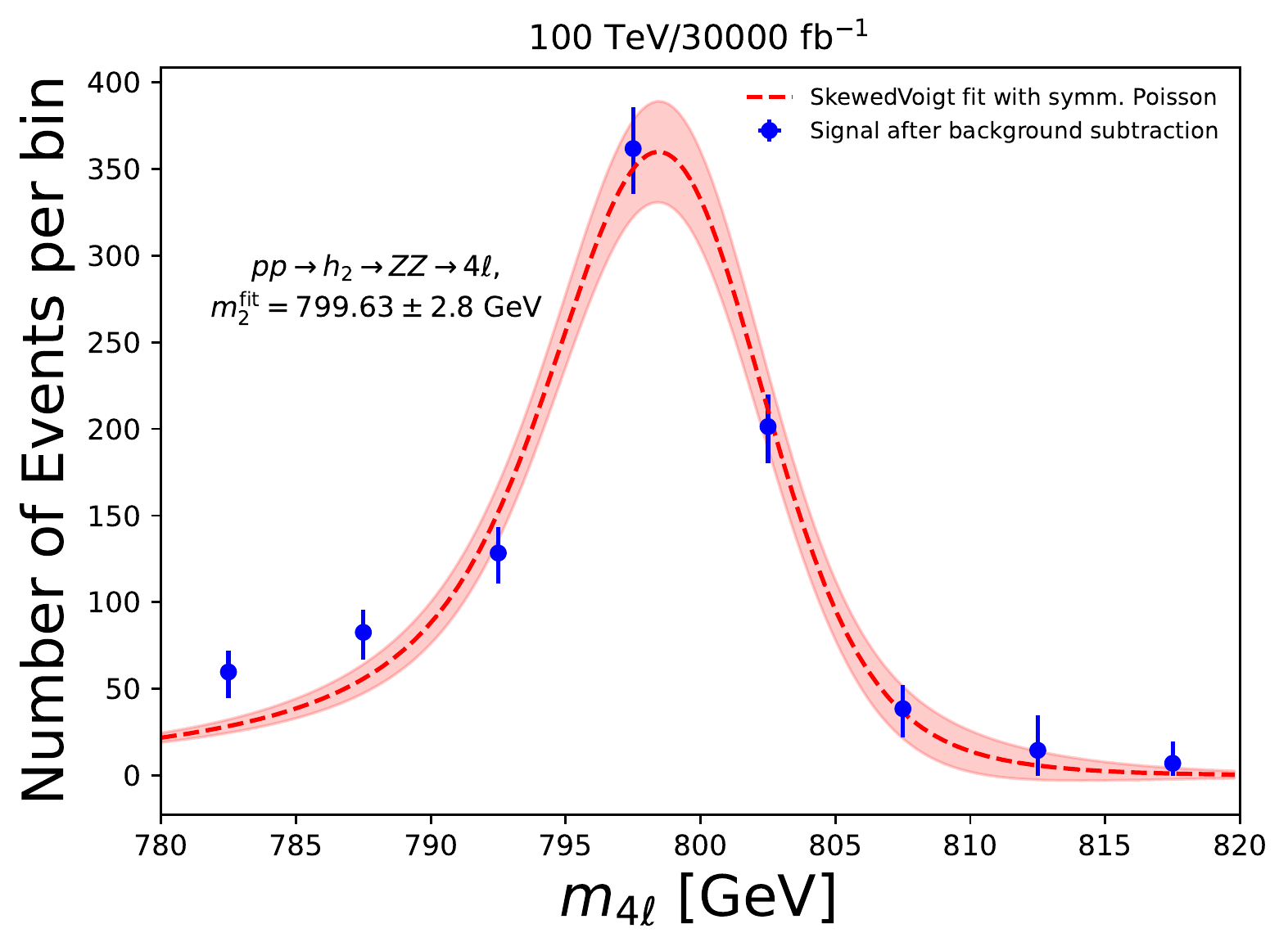}
  \includegraphics[width=0.45\columnwidth]{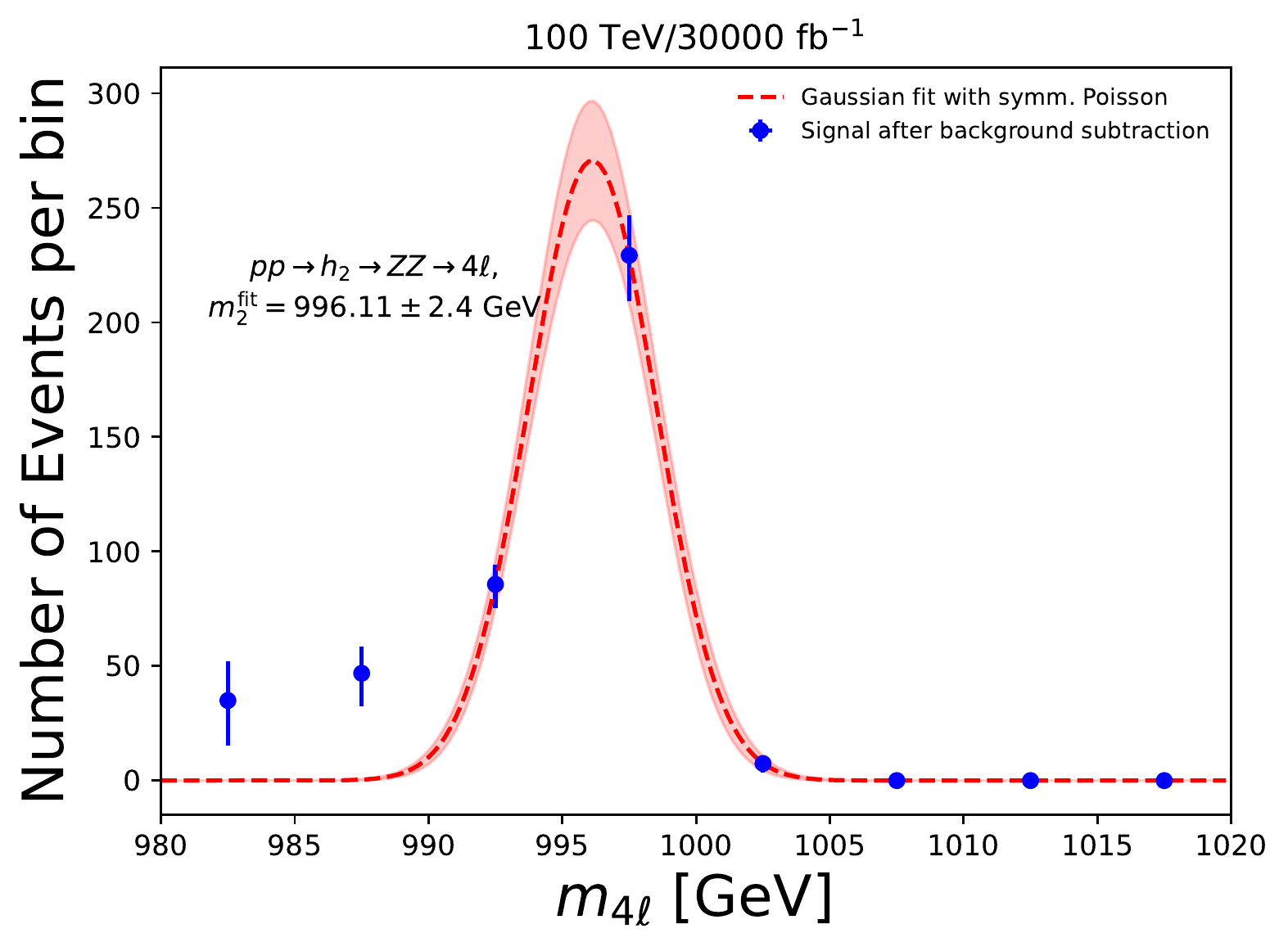}
\caption{A selection of results obtained through the mass fitting procedure outlined in the main text is shown. We show the fits for cross sections corresponding to significances of $\Sigma \simeq 10$ in the $pp\rightarrow h_2 \rightarrow ZZ \rightarrow 4\ell$ channel at a 100 TeV collider with an integrated luminosity of 30~ab$^{-1}$, for real masses $m_2=400, 600, 800, 1000$~GeV corresponding to top left, top right, lower left and lower right. The blue error bars represent the expected number of signal events in each bin and the red dashed curve and red error band represent the fit and its corresponding uncertainty as obtained by the \texttt{lmfit} package.}
\label{fig:massfitexamples_ZZ4l}
\end{figure}

\begin{figure}[htp]
  \centering
  \includegraphics[width=0.49\columnwidth]{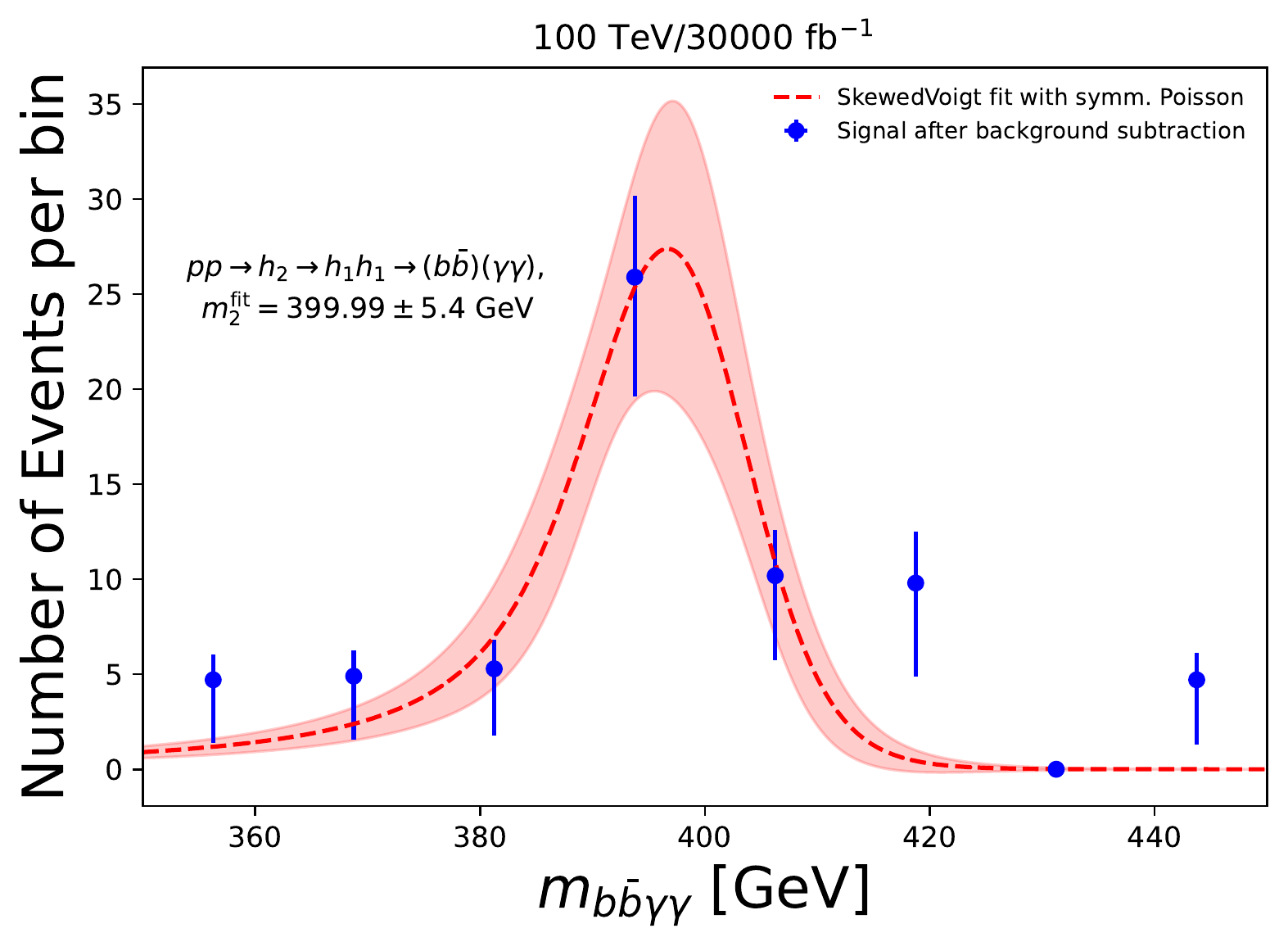}
  \includegraphics[width=0.49\columnwidth]{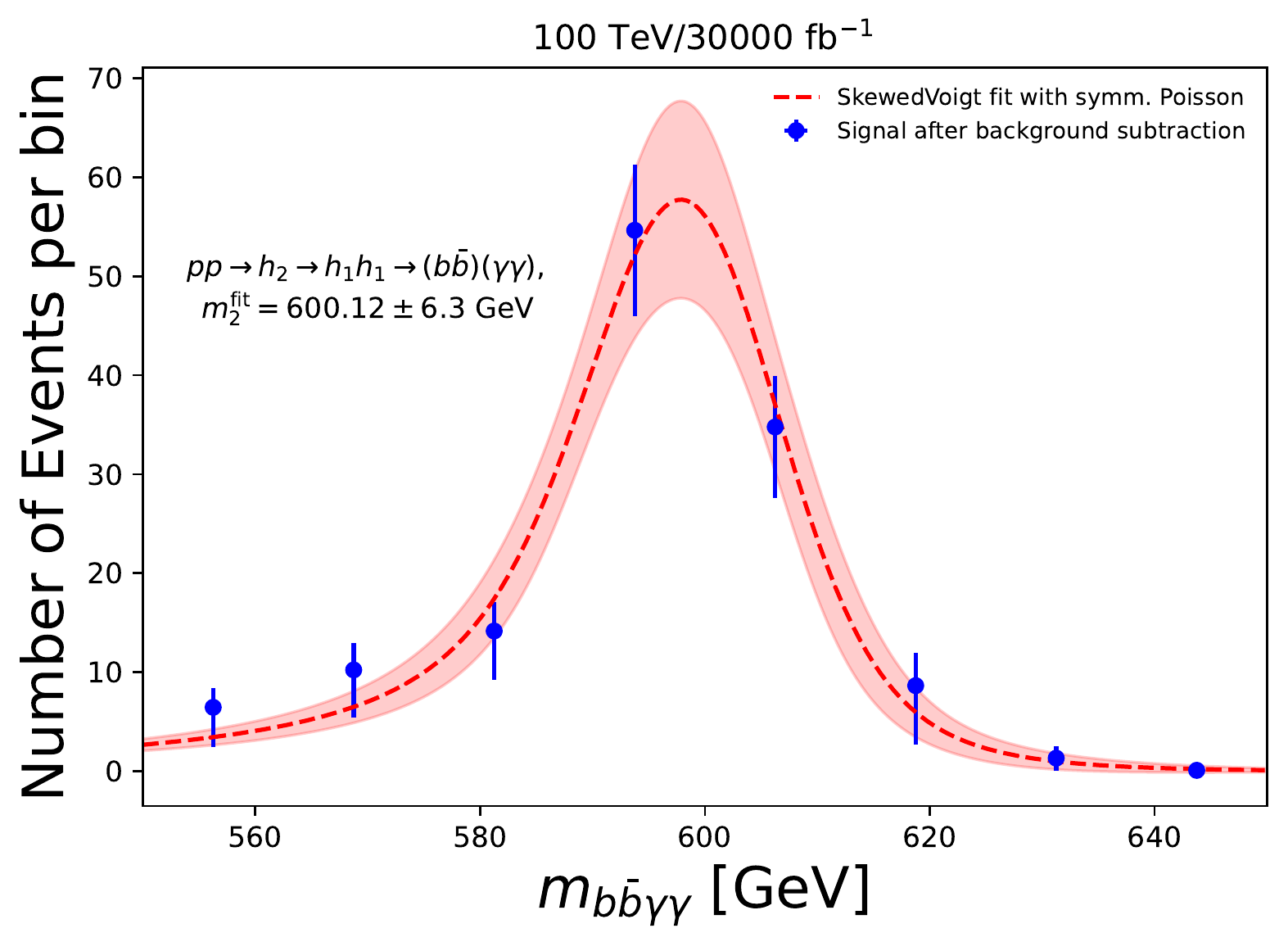}
  \includegraphics[width=0.49\columnwidth]{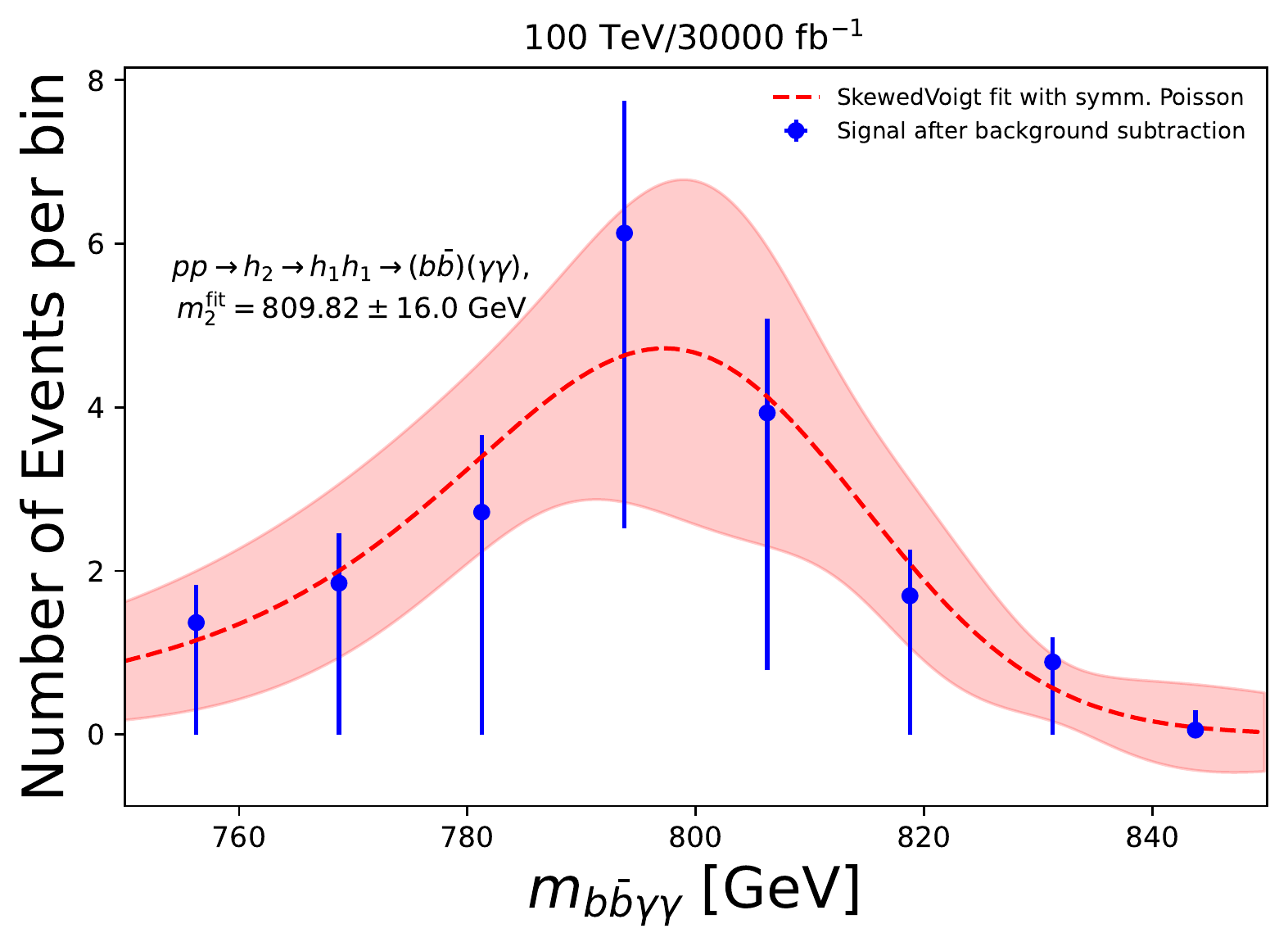}
  \includegraphics[width=0.49\columnwidth]{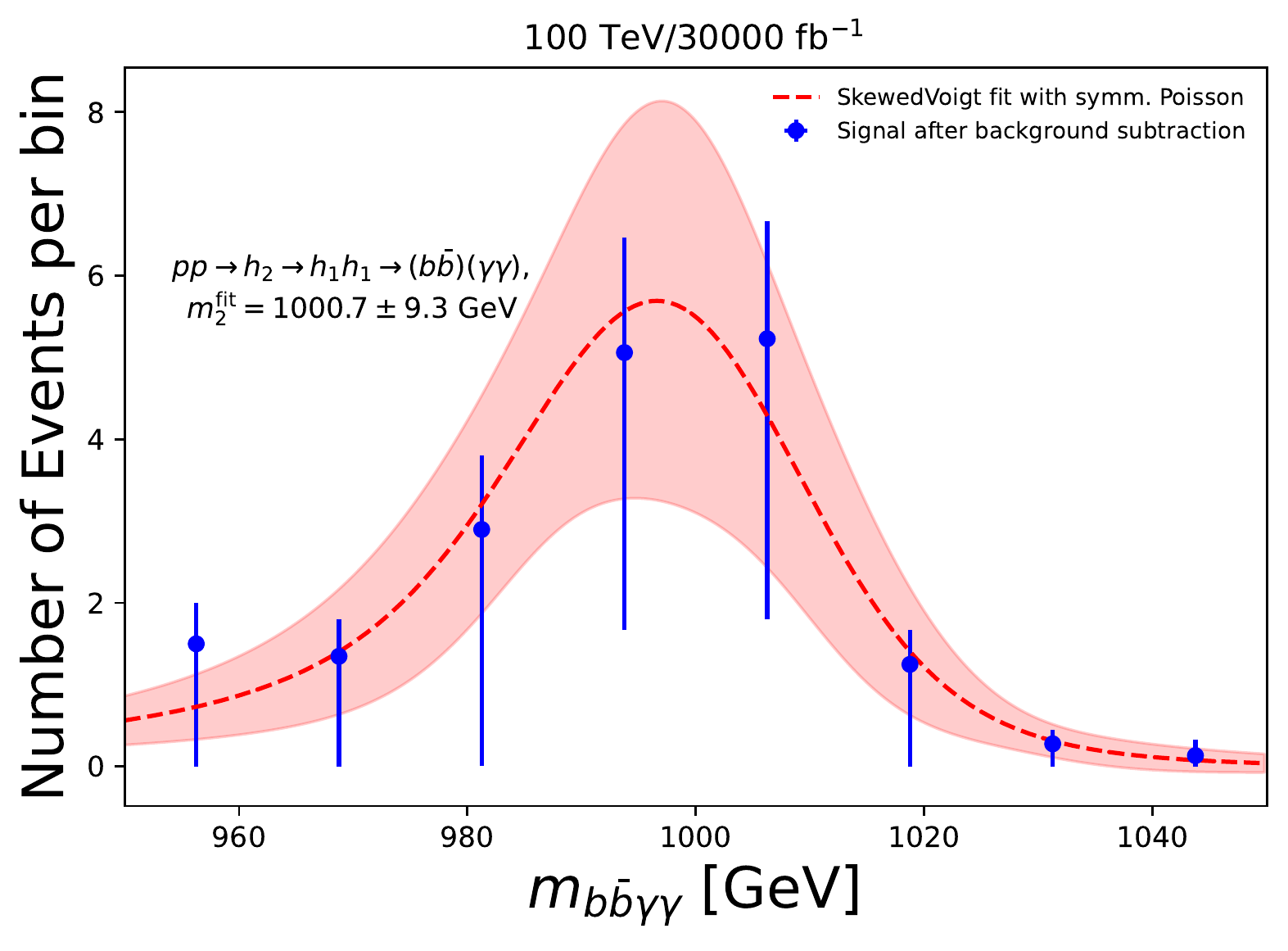}
\caption{A selection of results obtained through the mass fitting procedure outlined in the main text is shown. We show the fits for cross sections corresponding to significances of $\Sigma \simeq 20$ in the $pp\rightarrow h_2 \rightarrow h_1 h_1 \rightarrow (b\bar{b})(\gamma\gamma)$ channel at a 100 TeV collider with an integrated luminosity of 30~ab$^{-1}$, for real masses $m_2=400, 600, 800, 1000$~GeV corresponding to top left, top right, lower left and lower right. The blue error bars represent the expected number of signal events in each bin and the red dashed curve and red error band represent the fit and its corresponding uncertainty as obtained by the \texttt{lmfit} package.}
\label{fig:massfitexamples_HH}
\end{figure}

In Fig.~\ref{fig:massfitexamples_ZZ4l} we show a selection of fit results obtained through the outlined procedure through the four-lepton final state, for real masses $m_2=400, 600, 800, 1000$~GeV, corresponding to cross sections that yield significances of $\Sigma \simeq 10$ in the $pp\rightarrow h_2 \rightarrow ZZ \rightarrow 4\ell$ channel at a 100 TeV collider, with an integrated luminosity of 30~ab$^{-1}$. In Fig.~\ref{fig:massfitexamples_HH} we show the fits obtained for real masses $m_2=400, 600, 800, 1000$~GeV through the $pp\rightarrow h_2 \rightarrow h_1 h_1 \rightarrow (b\bar{b})(\gamma\gamma)$ channel, for a significance of $\Sigma \simeq 20$. The plots show the fit in red-dashed lines and the associated uncertainty in the red band, obtained for the blue error bars,\footnote{The error bars were symmetrised for the purposes of the fit.} which represent the expected signal after background subtraction. It is evident that the fits obtained through the four-lepton final state are expected to perform better than those obtained through the $h_2 \rightarrow h_1 h_1 \rightarrow (b\bar{b})(\gamma\gamma)$ process. 

\subsubsection{\boldmath Mass measurement through the transverse mass, $m_T$}

\begin{figure}[htp]
  \centering
  \includegraphics[width=0.49\columnwidth]{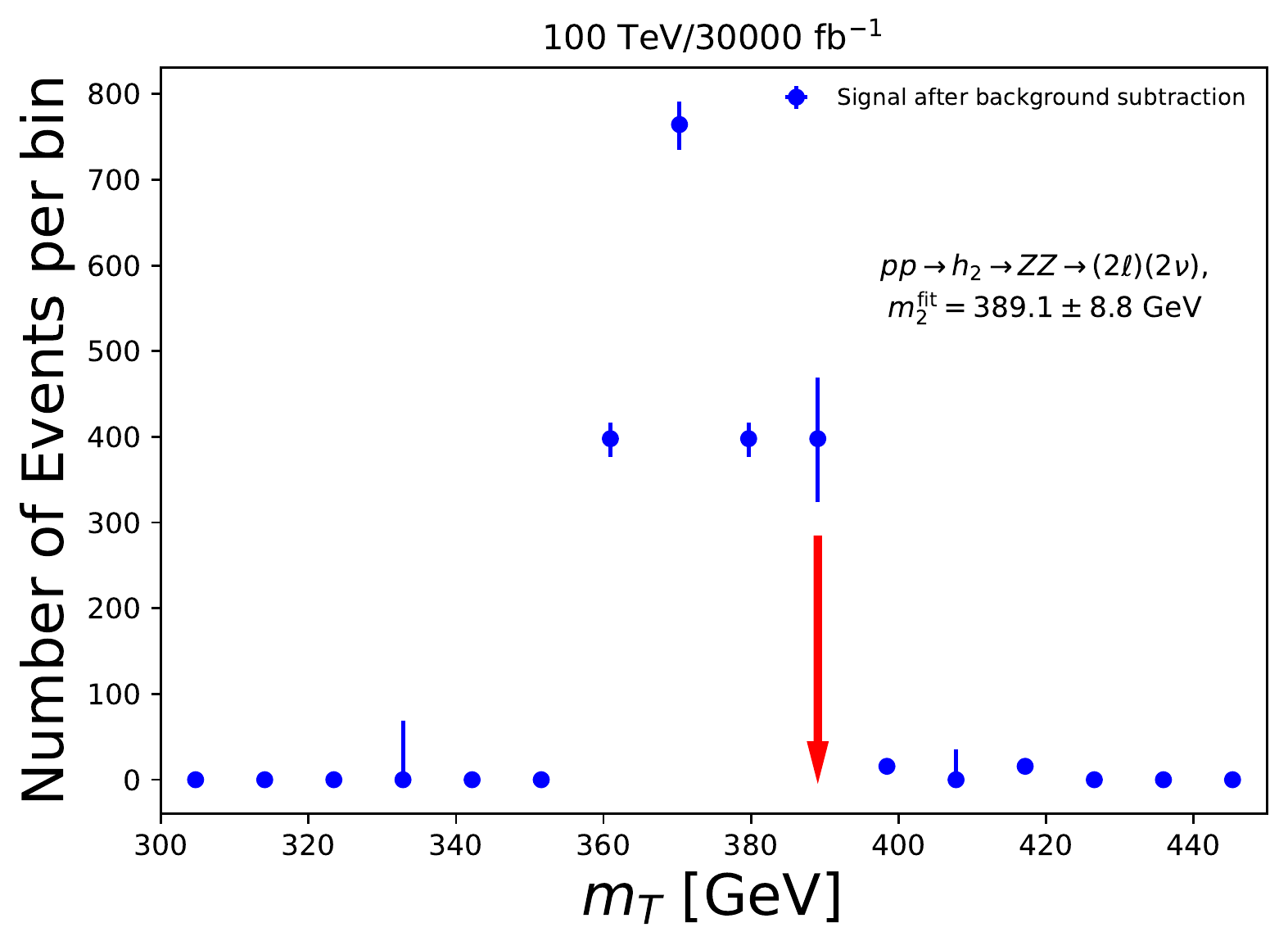}
  \includegraphics[width=0.49\columnwidth]{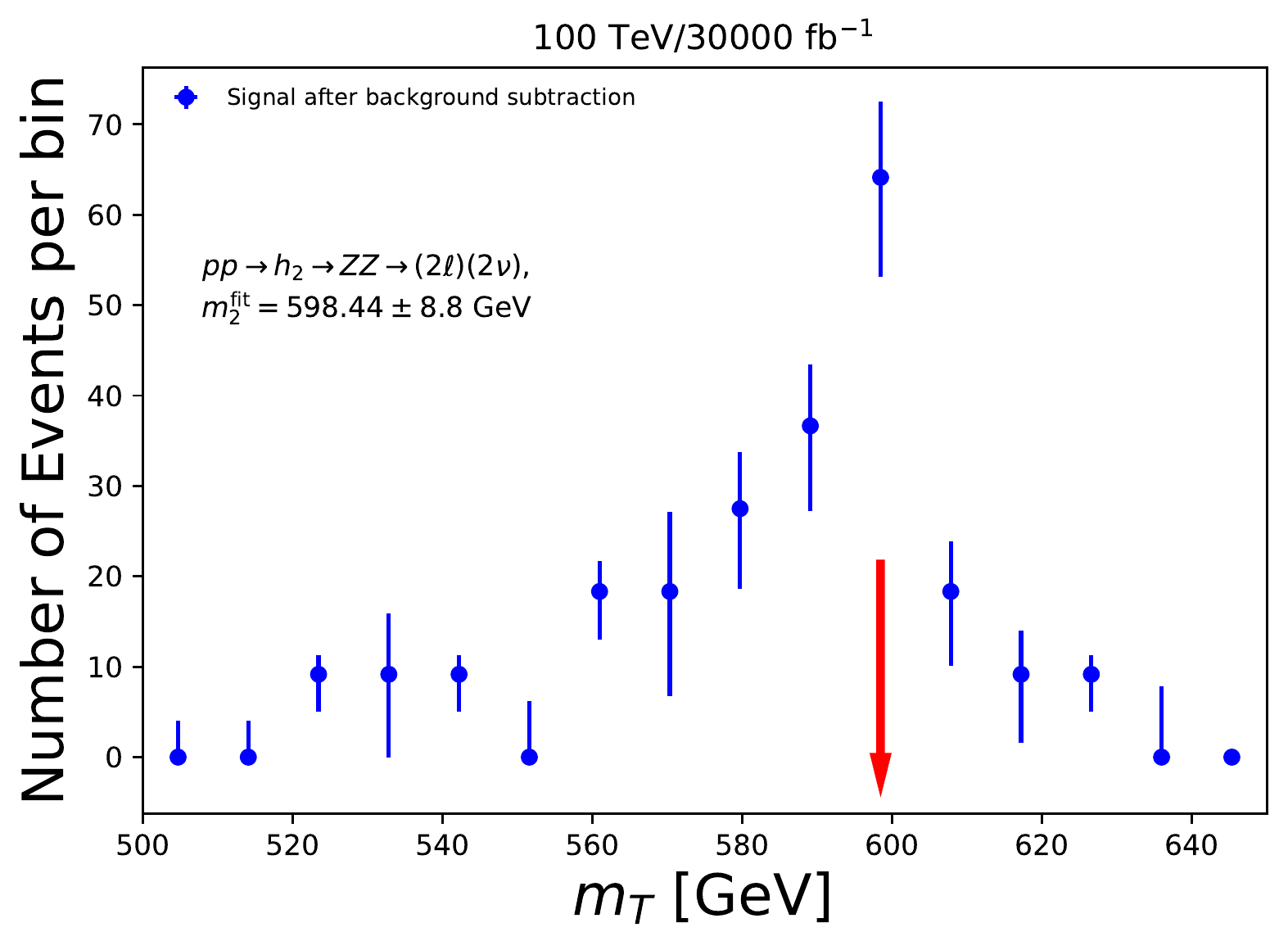}
  \includegraphics[width=0.49\columnwidth]{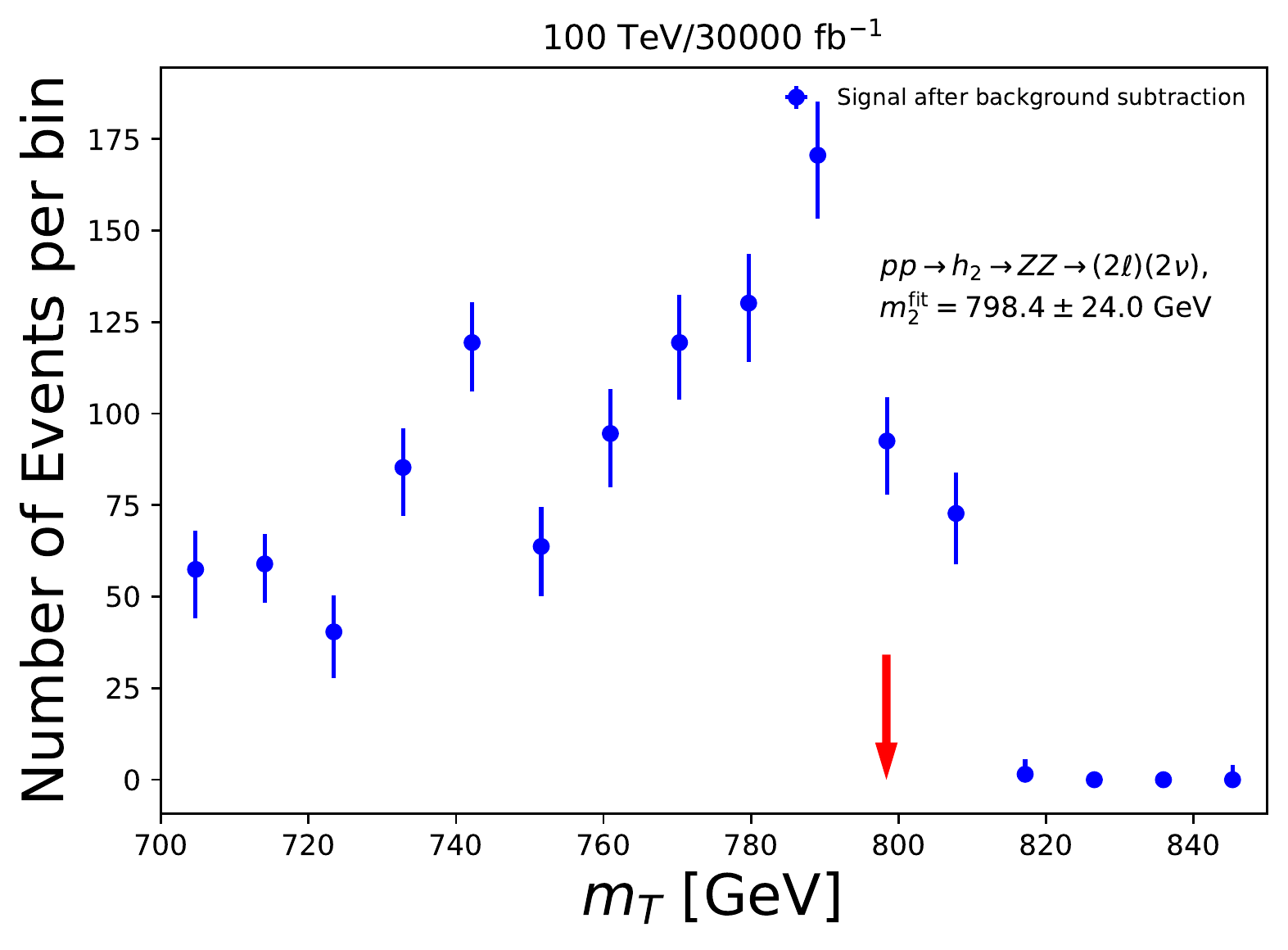}
  \includegraphics[width=0.49\columnwidth]{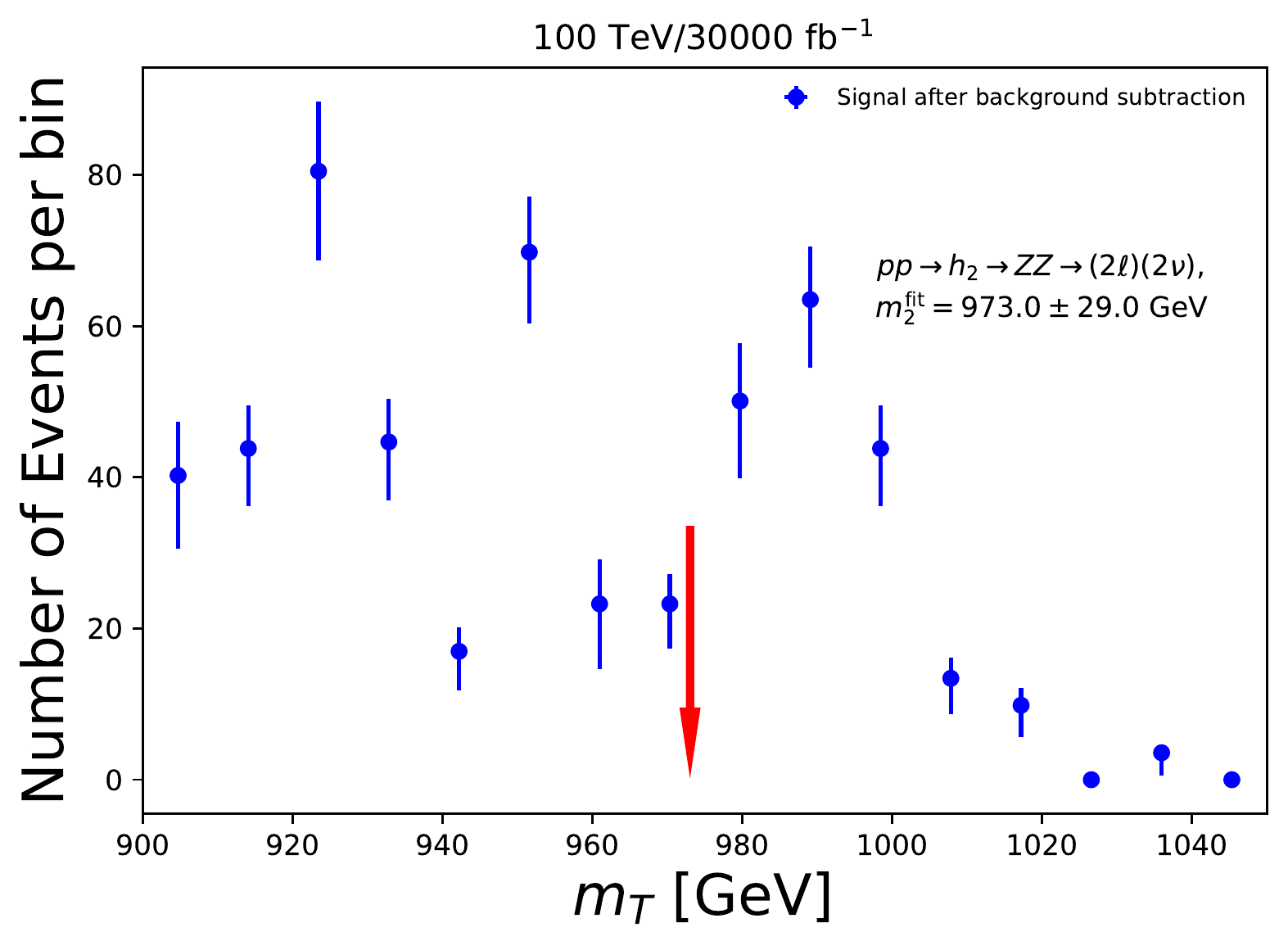}
\caption{A selection of results obtained through the mass determination procedure outlined in the main text is shown. We show the fits for cross sections corresponding to significances of $\Sigma \simeq 10$ in the $pp \rightarrow ZZ \rightarrow (\ell^+ \ell^-) (\nu \nu)$ channel at a 100 TeV collider with an integrated luminosity of 30~ab$^{-1}$, for real masses $m_2=400, 600, 800, 1000$~GeV corresponding to top left, top right, lower left and lower right. The blue error bars represent the expected number of signal events in each bin. The red arrow shoes the determined value of the mass $m_2$. The uncertainty on the resulting mass fit is determined by a combination of the bin width and the statistical error.}
\label{fig:massfitexamples_2l2v}
\end{figure}

The $pp \rightarrow ZZ \rightarrow (\ell^+ \ell^-) (\nu \nu)$ final state was found to be the most constraining in ref.~\cite{Papaefstathiou:2020iag}. However, due to the undetected neutrinos, it is not possible to reconstruct the invariant mass of the final-state objects and obtain a precise fit of the mass of the $h_2$. Nevertheless, the transverse mass variable, $m_T$, employed in the analysis and defined in eq.~\ref{eq:mt2l2v}, should provide an `edge' near the real mass of the $h_2$ since $m_T \leq m_2$. Modelling the shape of $m_T$, particularly after the analysis cuts and background subtraction are applied, is a complex task and therefore we have devised a strategy that yields a strong correlation between the mass and the `edge' of the distribution. In particular, we define the edge of the $m_T$ distribution as being represented by the largest `drop' in the number of events between two successive bins. The position of the edge is directly correlated with the mass $m_2$. Therefore, when estimating the mass, we consider the lower of the two bins that form the ratio to be the value of $m_2$. Since this method is susceptible to statistical fluctuations, to obtain an estimate of the statistical error on the position of the edge, we perform 400 pseudo-experiments for the expected number of events in each bin and calculate the mean and standard deviation of the edge position. To calculate the total uncertainty, we combine this statistical error with the bin width, which should represent an estimate of the ``systematic'' uncertainty on the mass measurement obtained through this technique. In Fig.~\ref{fig:massfitexamples_2l2v} we show the fits for cross sections corresponding to significances of $\Sigma \simeq 10$ obtained via this method, with an integrated luminosity of 30~ab$^{-1}$, for real masses $m_2=400, 600, 800, 1000$~GeV corresponding to top left, top right, lower left and lower right panels. The blue error bars represent the expected number of signal events in each bin and the red arrow indicates the position of the determined value of the mass $m_2$.

\subsection{\boldmath Measuring $\theta$ and $\lambda_{112}$}\label{sec:thetalambdafit}

The processes $pp \rightarrow h_2 \rightarrow h_1 h_1$ and $pp \rightarrow h_2 \rightarrow Z Z$ depend on the mixing angle $\theta$ and the $h_2 - h_1 - h_1$ coupling, $\lambda_{112}$. In the narrow-width approximation we may write the cross section as a product: $\sigma (pp \rightarrow h_2 \rightarrow yy) = \sigma(pp \rightarrow h_2) \times \mathrm{BR}(h_2 \rightarrow yy)$, where $yy = {h_1 h_1,~ZZ,~WW, ...}$ and $\mathrm{BR}(h_2 \rightarrow yy)$ is the corresponding branching ratio (BR) of $h_2$ to the $yy$ final state. The BRs for $h_2 \rightarrow xx$ (where $xx \neq h_1 h_1)$ and $h_2 \rightarrow h_1 h_1$ are then given by, respectively,
\begin{eqnarray}
    \mathrm{BR}(h_2 \rightarrow xx) &=& \frac{\sin^2 \theta~ \Gamma^\mathrm{SM} (h_2 \rightarrow xx)}{\Gamma(h_2)} \;;\;\; x\neq h_1\;,\\
    \mathrm{BR}(h_2 \rightarrow h_1 h_1)& =& \frac{ \Gamma (h_2 \rightarrow h_1 h_1) }{\Gamma(h_2)}\;,
\end{eqnarray}
where the total width of $h_2$ is given by
\begin{equation}
    \Gamma(h_2) = \sin^2 \theta \sum_{x\neq h_1} \Gamma^\mathrm{SM} (h_2 \rightarrow xx) + \Gamma (h_2 \rightarrow h_1 h_1)\;,
\end{equation}
and $\Gamma^\mathrm{SM} (h_2 \rightarrow xx)$ is the corresponding width of a scalar of mass $m_2$ to the SM final state $xx \neq h_1 h_1$ and the width $\Gamma (h_2 \rightarrow h_1 h_1)$ is given at tree level by:
\begin{equation}
\Gamma (h_2 \rightarrow h_1 h_1) = \frac{ \lambda^2_{112} \sqrt{1 - 4m_1^2 / m^2_2}}{8 \pi m_2}\;.
\end{equation}
The above equations imply that \textit{all} the BRs of the $h_2$ depend on both $\sin \theta$ and $\lambda_{112}$. In particular, the $h_2 \rightarrow xx~ (xx \neq h_1 h_1)$ processes depend on $\lambda_{112}$ through the total width. Therefore, to obtain precise measurements of $\sin \theta$ and $\lambda_{112}$, a combination of two or more final states is required. Note that at tree level, $\lambda_{112} \rightarrow 0$ as $\theta \rightarrow 0$, with the inverse not being true, i.e.\ there exist points with small $\lambda_{112}$ but non-zero $\theta$.

\begin{figure}[htp]
  \centering
  \includegraphics[width=0.49\columnwidth]{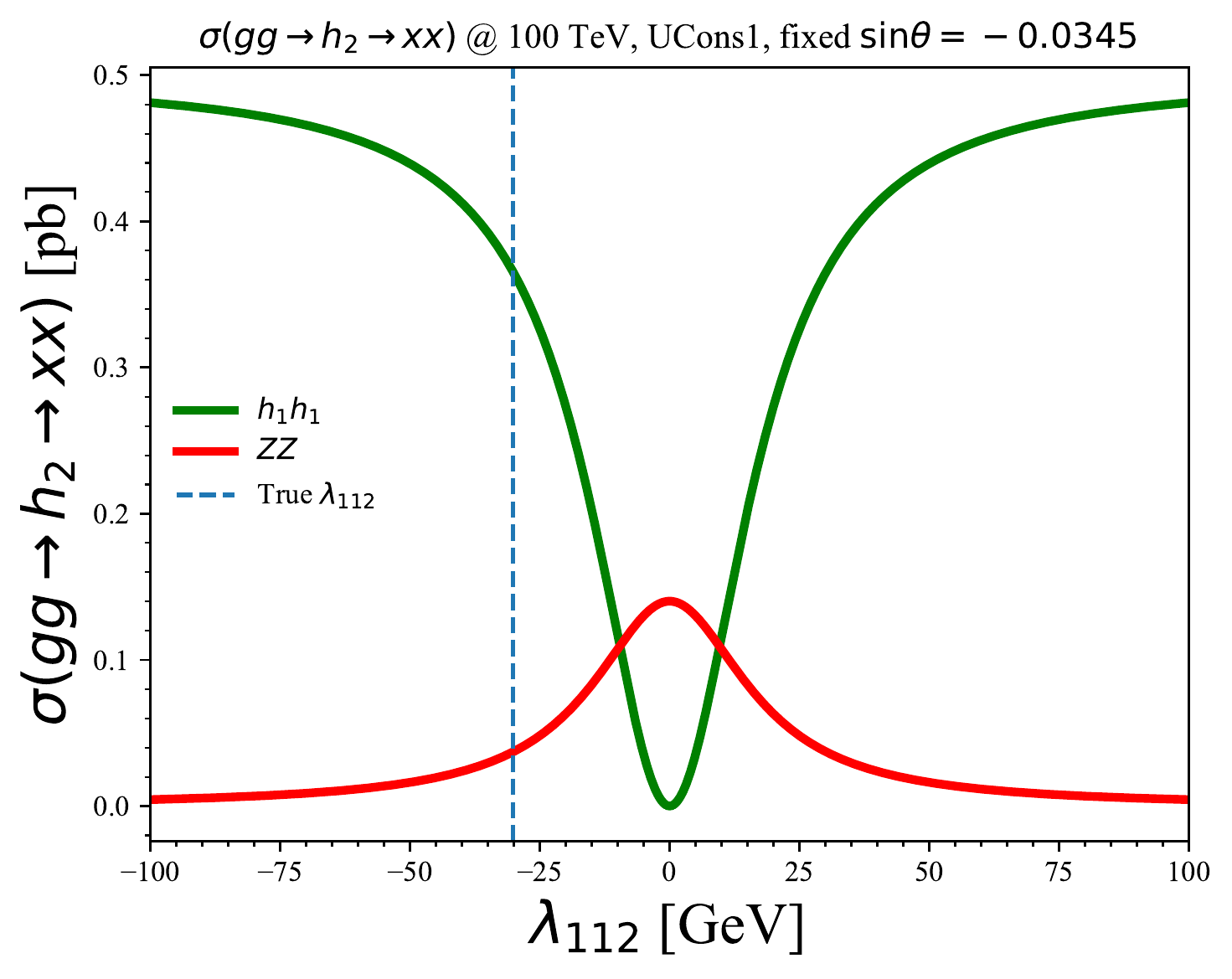}
  \includegraphics[width=0.49\columnwidth]{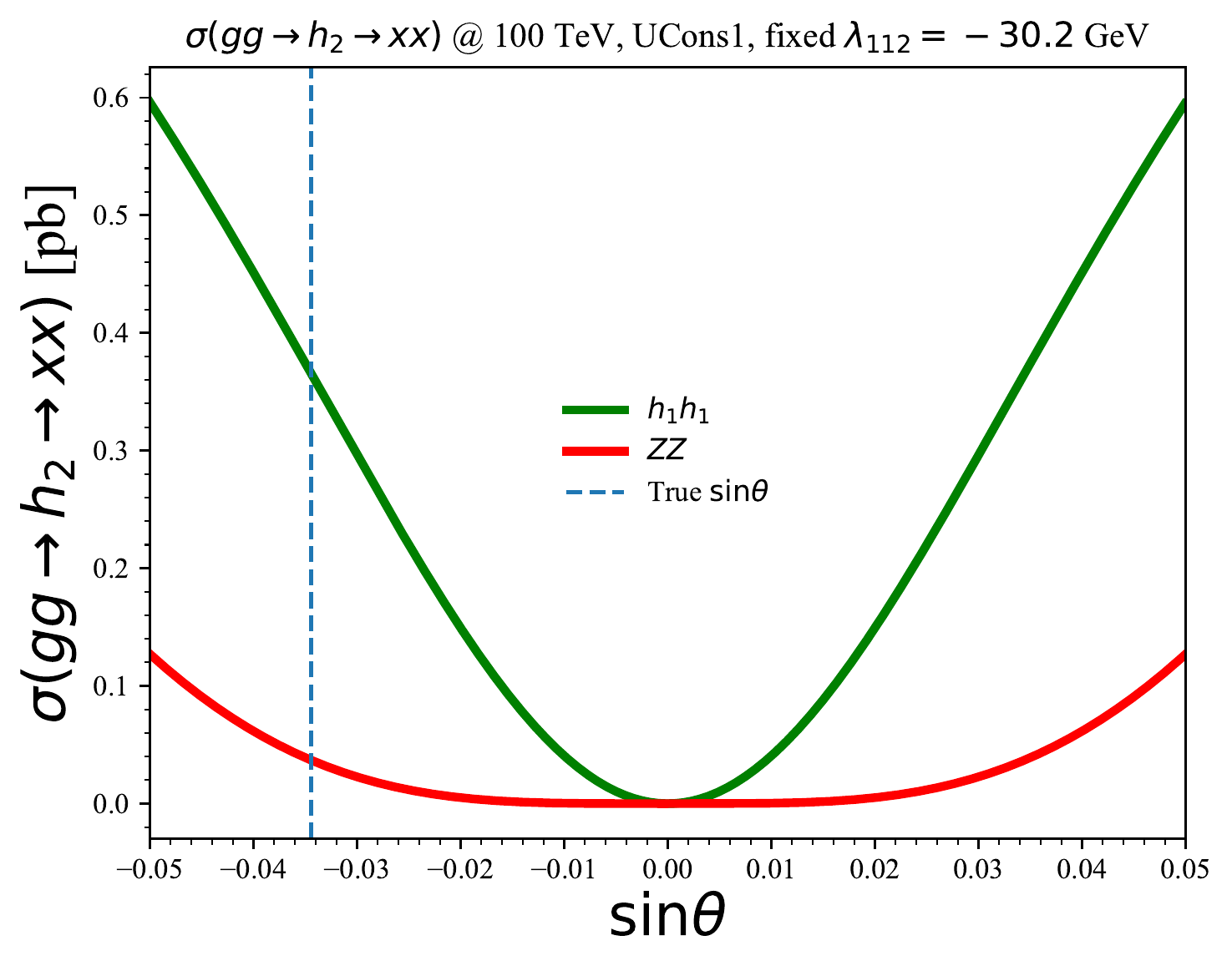}
\caption{The cross sections for  $pp \rightarrow h_2 \rightarrow h_1 h_1$ and $pp \rightarrow h_2 \rightarrow ZZ$ for the ``UCons1'' benchmark point of ref.~\cite{Papaefstathiou:2020iag}, as functions of either $\lambda_{112}$ (left) or $\sin \theta$ (right), keeping either $\sin \theta$ or $\lambda_{112}$ fixed to the true values, respectively.}
\label{fig:sigmavlambdaexample}
\end{figure}

\begin{table}[h]
    \centering
    \begin{tabular}{|l|c|}
    \hline
    \multicolumn{2}{|c|}{``UCons1'' Parameters} \\ \hline
    $\mathbold{\mu^2}$ \textbf{[GeV}$\mathbold{^{2}}$\textbf{]}& -2204.7 \\\hline 
    $\mathbold{M_S^2}$ \textbf{[GeV}$\mathbold{^{2}}$\textbf{]}& -8129.0  \\\hline 
    $\mathbold{K_1}$ \textbf{[GeV]}& -204.3  \\\hline 
    $\mathbold{K_2}$ & 4.33  \\\hline 
    $\mathbold{\kappa}$ \textbf{[GeV]}& -61.0 \\\hline \hline
    \textbf{sin}$\mathbold{\theta}$ &  -0.034 \\ \hline
    $\mathbold{m_2}$ \textbf{[GeV]} &  377.5 \\\hline
    $\mathbold{\lambda_{112}}$  \textbf{[GeV]} & -30.183 \\ \hline
    \end{tabular}
    \caption{The real singlet-extended SM potential parameters, the sine of the mixing angle, mass $m_2$ and the tree-level triple coupling $\lambda_{112}$, for the benchmark point ``UCons1'' as it appears in Table~2 of ref.~\cite{Papaefstathiou:2020iag}, used as an example for fitting here.}
    \label{tb:ucons1}
    \end{table}
    
In what follows, we have employed the $pp \rightarrow h_2 \rightarrow h_1 h_1$ and the $pp \rightarrow h_2 \rightarrow ZZ$ final states to extract a combined expected limit on $\sin \theta$ and $\lambda_{112}$ at a 100 TeV proton collider with an integrated luminosity of 30~ab$^{-1}$. We require that $m_2 > 260$~GeV, sufficiently above the threshold for $pp \rightarrow h_2 \rightarrow h_1 h_1$ to be active. In Fig.~\ref{fig:sigmavlambdaexample} we show an example of the cross sections for $pp \rightarrow h_2 \rightarrow h_1 h_1$ and $pp \rightarrow h_2 \rightarrow ZZ$ as functions of either $\lambda_{112}$ (left panel) or $\sin \theta$ (right panel), keeping either $\sin \theta$ or $\lambda_{112}$ fixed to the true values, respectively, for the ``UCons1'' benchmark point as it appears in Table~2 of ref.~\cite{Papaefstathiou:2020iag}. The tree-level parameters for this specific benchmark point are given in Table~\ref{tb:ucons1}.

The $pp \rightarrow h_2 \rightarrow ZZ$ process was shown in ref.~\cite{Papaefstathiou:2020iag} to possess the highest significance for discovery of the $h_2$. To obtain the fit for a particular parameter-space point, we use the significances obtained in ref.~\cite{Papaefstathiou:2020iag} for each of these final states. The significance, $\Sigma$, can be used to obtain an estimate on the statistical uncertainty on the cross section measurement, $\Delta \sigma$ as:\footnote{See appendix~\ref{app:xsecuncert} for an explanation of the origin of this formula.}
\begin{equation}\label{eq:xsecuncert}
    \Delta \sigma(xx) = \frac{\sigma(xx)}{\Sigma(xx)} \;,
\end{equation}
where we use $xx$ as a shorthand for $pp \rightarrow h_2 \rightarrow xx$. 

We then construct bands that contain $\sigma(h_1 h_1) \pm \Delta \sigma(h_1 h_1)$ and $\sigma(ZZ) \pm \Delta \sigma(ZZ)$ over the $(\lambda_{112}, \sin \theta)$ plane. We will assume that the constraints on $(\lambda_{112}, \sin \theta)$ are represented by the overlap of the bands corresponding to the two processes. This should provide a conservative estimate of the one-standard deviation limits on the plane.\footnote{To do this more precisely, one should calculate the overlap between the $p$-value distributions within the bands. Given the other uncertainties in our process and for simplicity, we do not take this approach here.}

To incorporate the uncertainty in the mass measurement as described in section~\ref{sec:massfit}, $\Delta m_2^{\mathrm{fit}}$, we calculate the $\sigma(h_1 h_1)$ and $\sigma(ZZ)$ bands including the variation of the mass within one standard deviation, i.e.\ we calculate four bands: $\sigma(h_1 h_1; m_2\pm \Delta m_2^{\mathrm{fit}})$ and  $\sigma(ZZ; m_2\pm \Delta m_2^{\mathrm{fit}})$. We then take the largest parallelogram region obtained by the overlap of these four bands to represent the region of constraint of $(\lambda_{112}, \sin \theta)$. Note that since both of these processes depend on the squares of both $\lambda_{112}$ and $\sin \theta$, there will always be a sign ambiguity for any constraint obtained through their combination. 

\begin{figure}[htp]
  \centering
  \includegraphics[width=0.45\columnwidth]{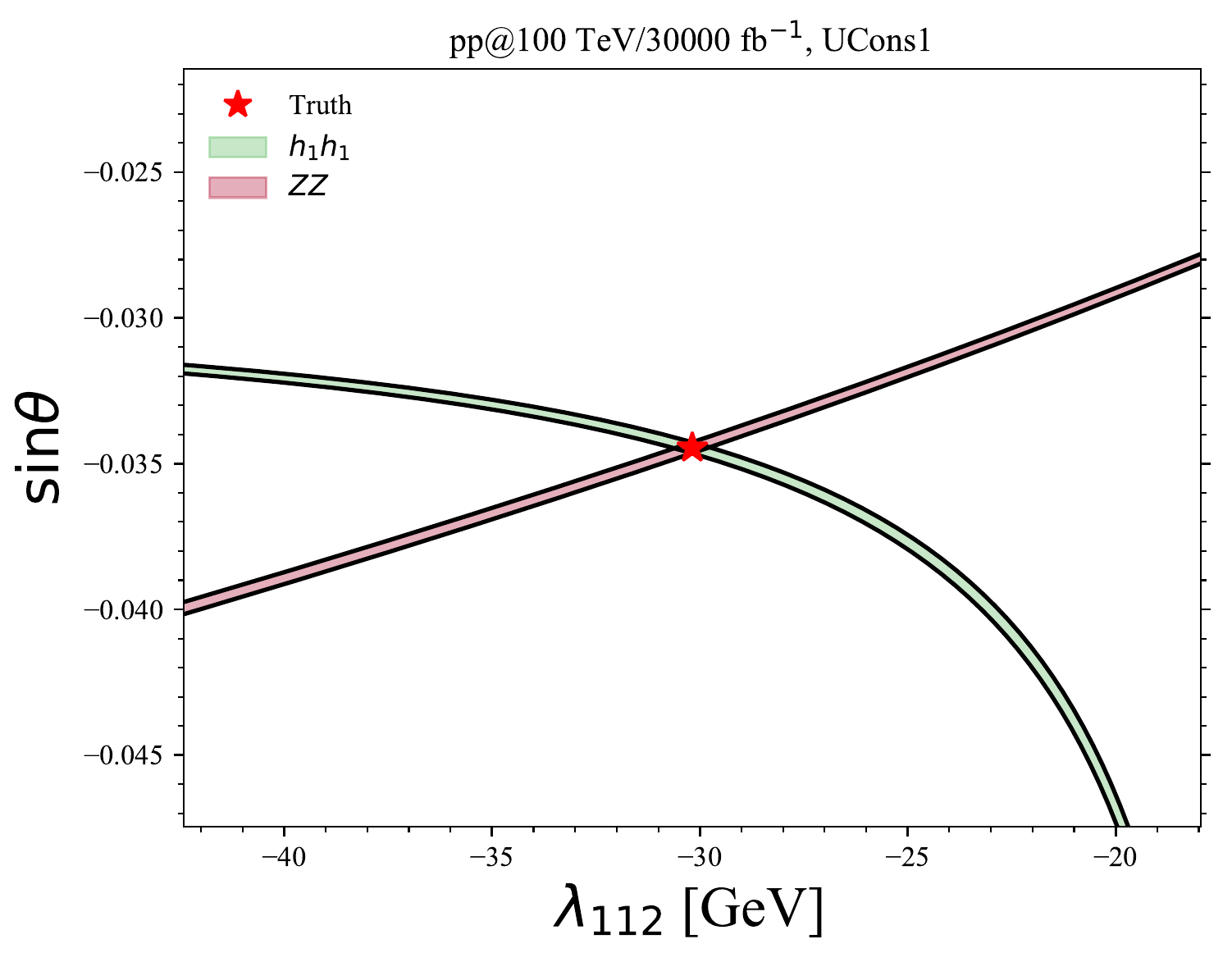}
  \includegraphics[width=0.45\columnwidth]{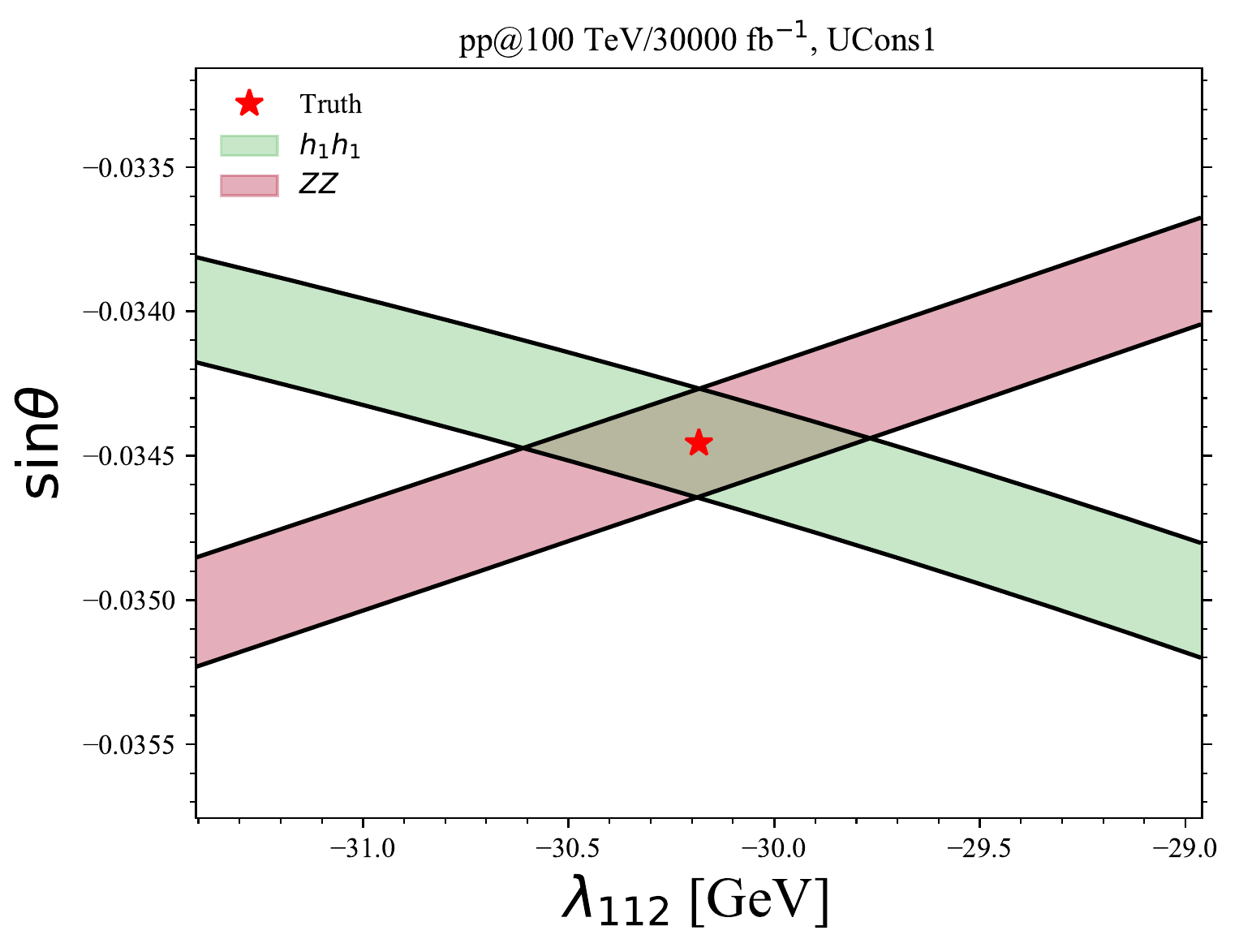}
\caption{An example of the fitting procedure for $(\lambda_{112}, \sin \theta)$ using the $\sigma(h_1 h_1)$ and $\sigma(ZZ)$ bands as described in the text. The fit was again performed for the benchmark point ``UCons1''. The red star represents the true values of $\sin \theta \approx -0.034$, $\lambda_{112} \approx -30.183$~GeV. The significances for this point in the $h_1 h_1$ and $ZZ$ final states from the analyses of ref.~\cite{Papaefstathiou:2020iag} was found to be $\Sigma(h_1 h_1) \approx 123.4$ and $\Sigma (ZZ) \approx 53.0$ standard deviations. This particular example was performed while keeping the mass fixed at the true value $m_2=377.5$~GeV, for simplicity. The left plot shows an enlarged region and the right plot is a zoomed-in version.}
\label{fig:lambdasinthetafitexample}
\end{figure}

We show an example of the fitting procedure for $(\lambda_{112}, \sin \theta)$ using the $\sigma(h_1 h_1)$ and $\sigma(ZZ)$ bands in Fig.~\ref{fig:lambdasinthetafitexample}. The fit was performed for the benchmark point ``UCons1''. The red star represents the true values of $\sin \theta \approx -0.034$, $\lambda_{112} \approx -30.183$~GeV. The significance for this point in the $h_1 h_1$ and $ZZ$ final states from the analyses of ref.~\cite{Papaefstathiou:2020iag} was found to be $\Sigma(h_1 h_1) \approx 123.4$ and $\Sigma (ZZ) \approx 53.0$ standard deviations. This particular example was performed while keeping the mass fixed at the true value $m_2=377.5$~GeV, for simplicity. The left plot in the figure shows an enlarged region and the right plot is a zoomed-in version. The constraints were found to be $|\lambda_{112}| \in [29.81, 30.65]$~GeV and $|\sin \theta| \in [ 0.0343, 0.0347]$, representing $\sim 3\%$ and $\sim 1\%$ precision, respectively, in line with the observed significances.

\section{Exploring the real singlet-extended SM parameter space}\label{sec:exploration}

\begin{figure}[htp]
  \centering
  \includegraphics[width=0.49\columnwidth]{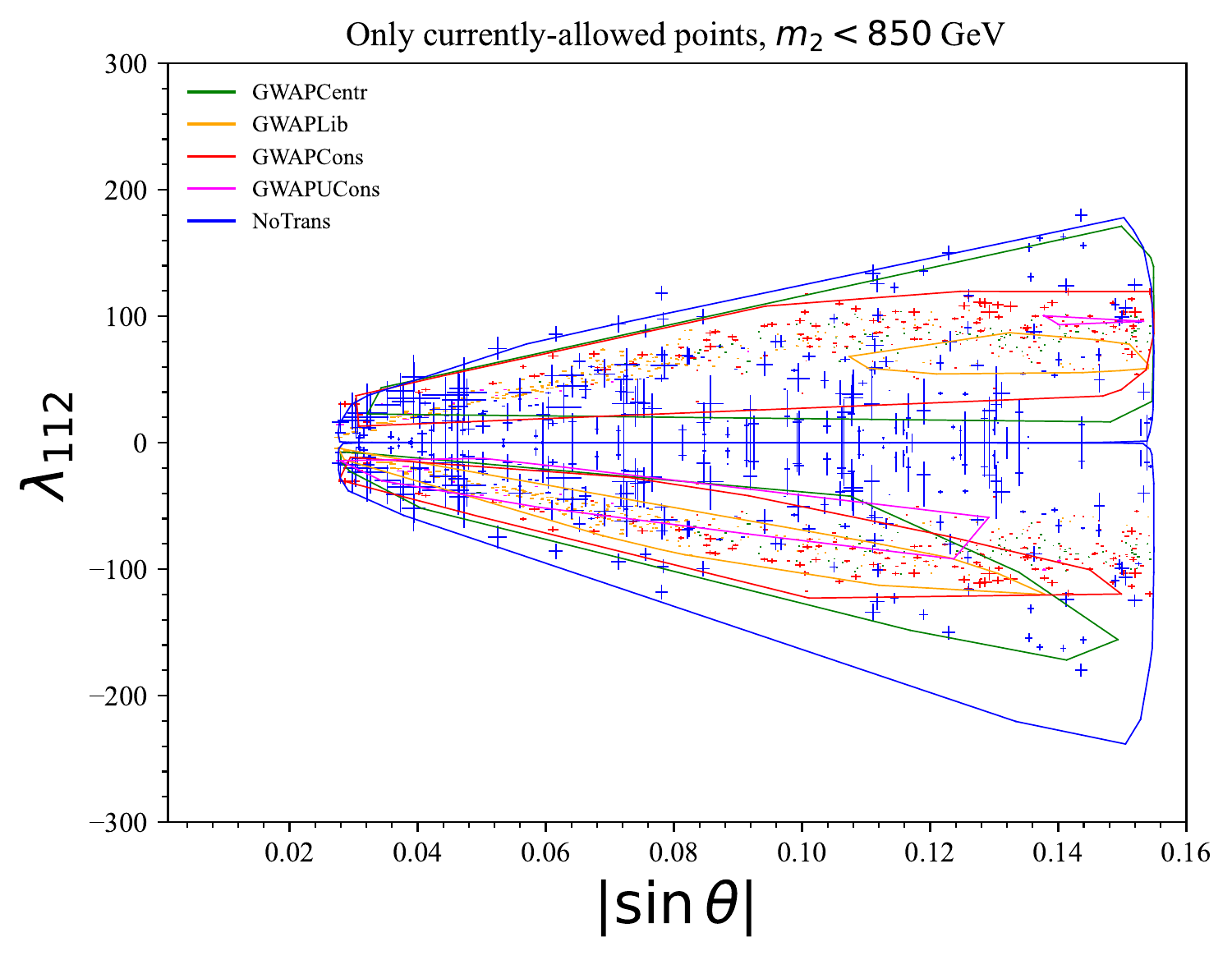}
  \includegraphics[width=0.49\columnwidth]{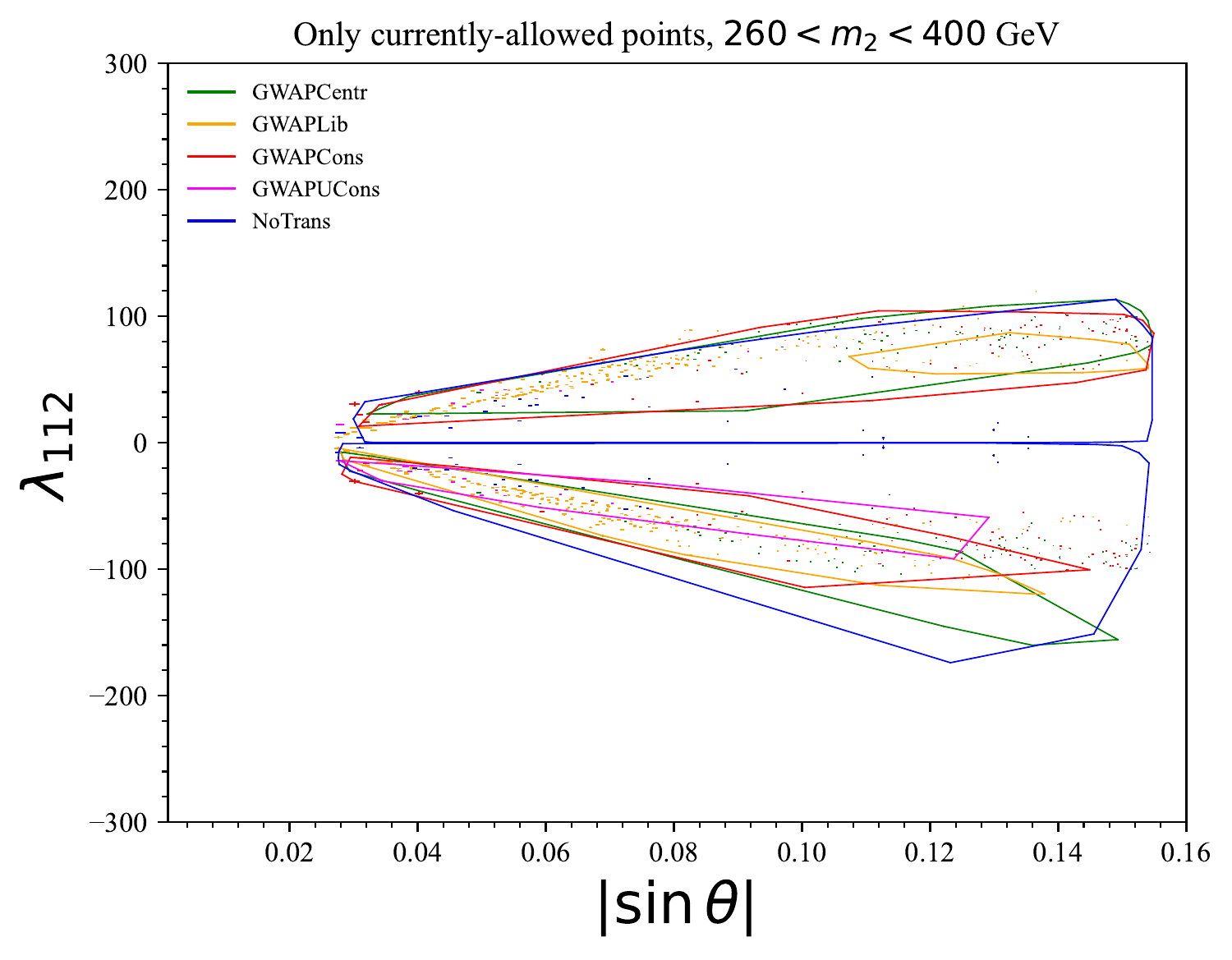}
  \includegraphics[width=0.49\columnwidth]{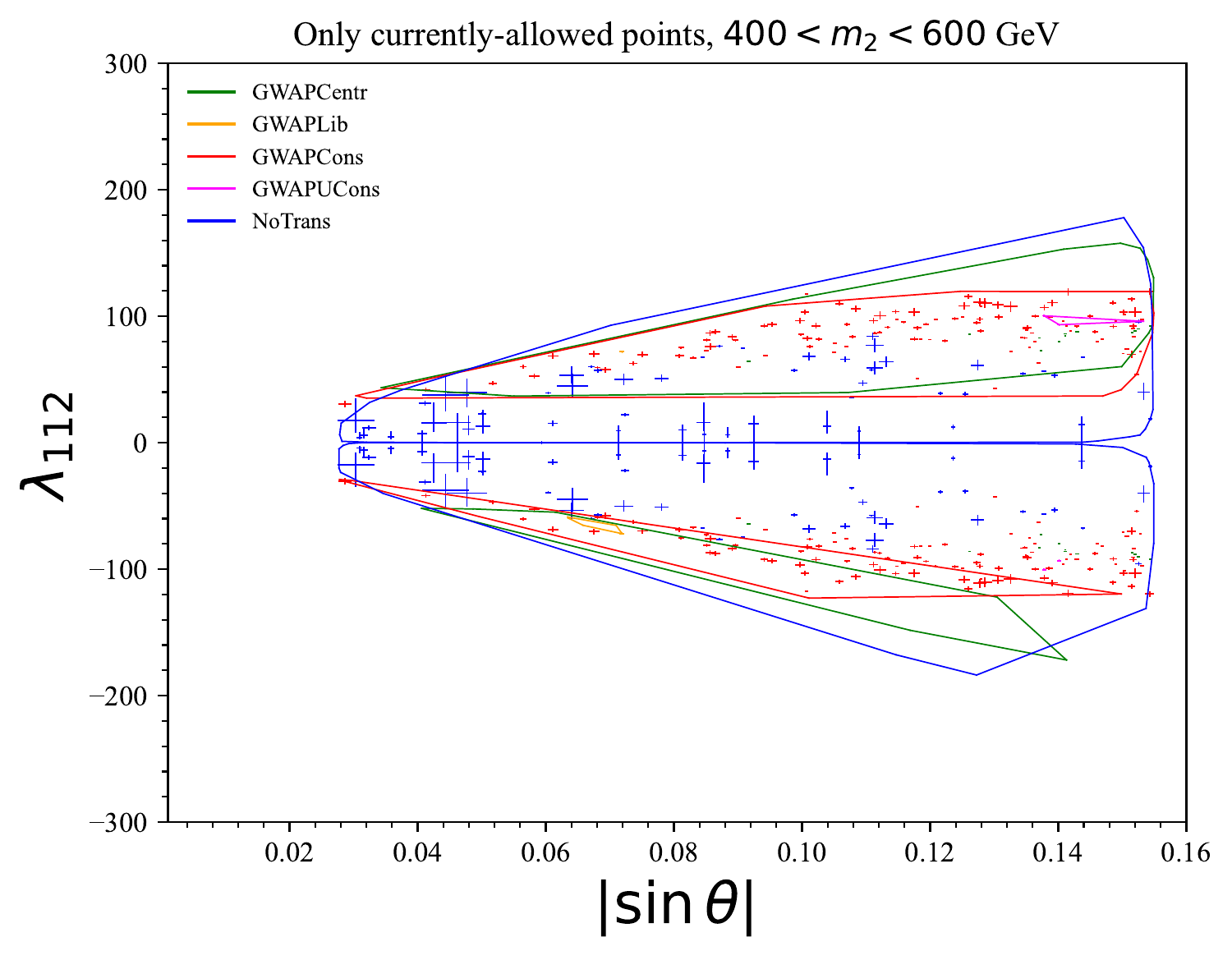}
  \includegraphics[width=0.49\columnwidth]{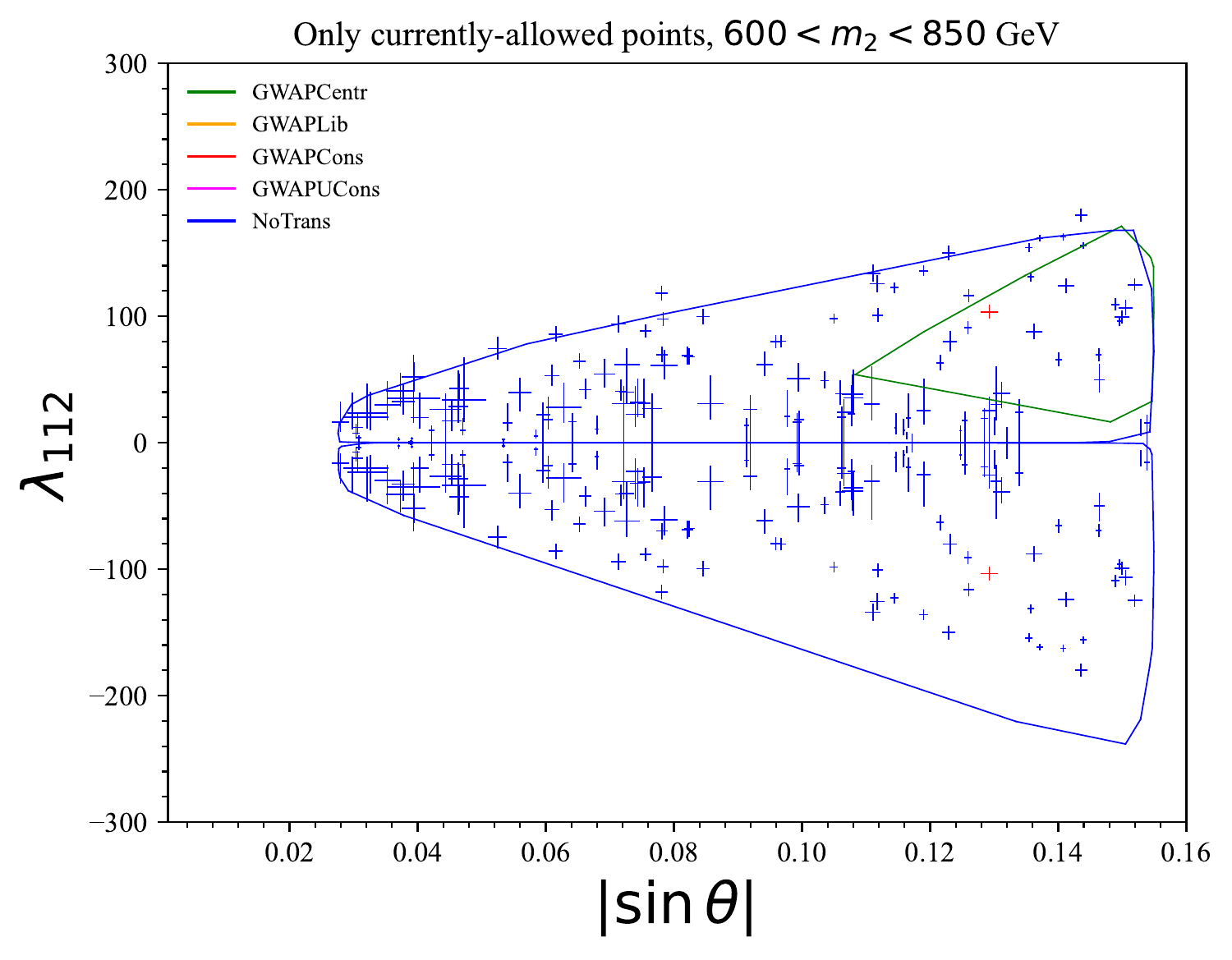}
\caption{The boundaries of the $(|\sin \theta|, \lambda_{112})$ parameter space for $m_2 < 850$~GeV (top left), $m_2 \in [260, 400]$~GeV (top right), $m_2 \in [400, 600]$~GeV (bottom left) and $m_2 \in [600, 850]$~GeV (bottom right). The error bars show the expected 1$\sigma$ constraints obtained through the procedure outlined in the article.}
\label{fig:sinthetal112}
\end{figure}

Putting everything together, we have performed a scan over the viable parameter space of the real singlet scalar extension of the SM. We have considered the ``Liberal'', ``Central'', ``Conservative'' and ``Ultra-Conservative'' points as discussed in section~\ref{sec:phasetrans} and defined in detail in ref.~\cite{Papaefstathiou:2020iag}. We re-iterate that the naming represents a decreasing degree of theoretical uncertainty (from ``Liberal'' to ``Conservative'') characterising how likely it is for a point in each category to generate a SFO-EWPT. In our scan, we also include ``NoTrans'' points, i.e.\ those that \textit{failed} to generate a SFO-EWPT during the parameter-space scan of ref.~\cite{Papaefstathiou:2020iag}, according to the set out criteria. The purpose of this exercise is to check whether the regions that exhibit SFO-EWPT and those that do not, are separated. If this is the case, the separation would allow us to say with certainty, following initial measurements, whether we can exclude or verify SFO-EWPT.

For a selection of benchmark points, we have calculated the expected 1$\sigma$ constraints on $|\sin \theta|$ and $\lambda_{112}$ according to the method described in section~\ref{sec:thetalambdafit}. The results of the scan are shown in Fig.~\ref{fig:sinthetal112}. On the top-left panel of the figure, we show the boundaries of the viable parameter space as defined by the points in this plane for each category. We have imposed the constraint $m_2 < 850$~GeV, since no points that generate SFO-EWPT were found beyond $m_2 \approx 850$~GeV. Note that there also exists a lower limit given by $m_2 \geq 260$~GeV such that $h_2 \rightarrow h_1 h_1$ is open and significant. The remaining three figures represent ``slices'' of the parameter space over the masses in the ranges $m_2 \in [260, 400]$~GeV (top right), $m_2 \in [400, 600]$~GeV (bottom left) and $m_2 \in [600, 850]$~GeV (bottom right). Several observations can be made, in the case of discovery of a new scalar particle:
\begin{itemize}
    \item For the mass bin $m_2 \in [600, 850]$~GeV, ``Centrist'' points appear to dominantly generate SFO-EWPT. In that case, if ``NoTrans'' points indeed represent the ``truth'', SFO-EWPT can be excluded for most of the parameter space except a small portion of the ``Centrist'' parameter space with $|\sin \theta| \in [\sim 0.10, \sim 0.15]$ and $\lambda_{112} \in [20, 150]$~GeV. This is despite the fact that measurement of the parameters of several ``NoTrans'' points is challenging, yielding large uncertainties in this mass bin. Only ``NoTrans'' points very near the ``Centrist'' region boundary may be mistaken for SFO-EWPT points.
    
    \item For the mass bins $m_2 \in [400, 600]$~GeV and $m_2 \in [260, 400]$~GeV, measurement of both $|\sin \theta|$ and $\lambda_{112}$ is expected to be very precise all over the viable parameter space. Therefore, if a point is measured to lie \textit{outside} the viable SFO-EWPT parameter-space boundaries, this will likely imply exclusion of SFO-EWPT to a high degree of certainty. 
    
    \item Irrespective of the value of $m_2$, a point is measured to lie \textit{within} the viable SFO-EWPT parameter space on the $(|\sin \theta|, \lambda_{112})$ plane, it will be challenging to exclude or verify SFO-EWPT. Therefore, additional measurements will become necessary. 

\end{itemize}

In the last case, the most likely complementary information can come from two additional sources: (i) Measurements of additional processes at colliders that contain multiple scalar $h_i$ bosons, the most promising of which would be the asymmetric process $pp\rightarrow h_1 h_2$ and\footnote{See, e.g.~\cite{Carmona:2016qgo} for a determination of a triple scalar coupling in a similar model.} (ii) measurements of gravitational waves. 

Multi-scalar processes would provide constraints on additional multi-scalar couplings, e.g. $pp\rightarrow h_1 h_2$ on the triple $h_1-h_2-h_2$ coupling $\lambda_{122}$, which would add a further dimension to the $(|\sin \theta|, \lambda_{112})$ plane of Fig.~\ref{fig:sinthetal112}. The constraints could allow us to map out additional regions where there exists a separation between the SFO-EWPT points and the ones that do not yield the right conditions for a SFO-EWPT. On the other hand, by measuring gravitational waves, on top of qualitatively giving information on the nature of the electro-weak transition, observing the peak frequency and peak amplitude could in principle fix two parameters which would then completely determine the parameter set.\footnote{Or perhaps more if the spectral shape can be determined and give information on the bubble wall velocity \cite{Gowling:2021gcy}.} Alternatively, a null observation could restrict the parameter space under the assumption that the reheating temperature was sufficiently high. The analysis of both of these avenues is left to future work. 

\section{Conclusions}\label{sec:conclusions}

The nature of electro-weak symmetry breaking is the one of the most fundamental questions facing next generation experiments. Such a question constitutes a key motivation for next-generation collider experiments and gravitational wave detectors. The nature of the transition may also provide clues to one of the most intriguing open questions -- that of the observed gigantic asymmetry between matter and anti-matter. Here we examined, for the first time, whether first measurements following the discovery of a new scalar particle at a future collider experiments alone could definitively uncover the nature of the transition and argued that this is \textit{not} the case, despite the fact that interesting and important information can be obtained. 

To go beyond the initial ``post-mortem'' measurements available upon the discovery of a new heavy scalar particle presented here, one can make use of additional, much rarer, processes at colliders, such as those that contain multiple Higgs bosons and/or new scalar bosons, e.g. $ pp \rightarrow h_1 h_2$, $ pp \rightarrow h_2 h_2$, $pp \rightarrow h_1 h_1 h_1$ and so on.\footnote{See e.g.~\cite{Papaefstathiou:2019ofh, Papaefstathiou:2020lyp} for studies of triple Higgs boson production in models with additional scalars. In addition, the multi-scalar processes could prove useful in the cases where $m_2 < 2 m_1$, where resonant $pp \rightarrow h_2 \rightarrow h_1 h_1$ is absent. For a study of rare Higgs boson decays in the case of $m_2 < m_1/2$, see~\cite{Kozaczuk:2019pet}. Finally, note that the final states that involve $h_2 h_2 + X$ might constitute a discovery channel in the so-called ``nightmare scenario'', where a $\mathbb{Z}_2$ symmetry is imposed, discussed in detail in~\cite{Curtin:2014jma}. In that case $pp \rightarrow h_1 h_2$ is absent and $pp \rightarrow h_1 h_1$ receives loop-level scalar contributions. The nightmare scenario is not covered by our study and deserves a full complementary investigation in its own right, which we leave to future work.} Furthermore, an additional source of information can arise from gravitational wave detectors, as even a null result could potentially constrain the parameter space significantly. Whether the combination of rare multi-scalar collider processes and gravitational wave detectors would allow for a full determination the nature of the electro-weak phase transition remains a significant open question that we leave to future endeavours.

\acknowledgments

We would like to thank Tuomas Tenkanen and Nikiforos Nikiforou for useful discussions. GW is supported by World Premier International Research Center Initiative (WPI), MEXT, Japan.

\appendix
\section{Phenomenological analyses at a 100 TeV proton collider}\label{app:analyses}

We outline here the main features of the analyses employed in the mass reconstruction of section~\ref{sec:massfit}. For the full details, including on additional final states, the event generation through  \texttt{MadGraph5\_aMC@NLO}~\cite{Alwall:2014hca, Hirschi:2015iia} and \texttt{HERWIG 7}~\cite{Bahr:2008pv, Gieseke:2011na, Arnold:2012fq, Bellm:2013hwb, Bellm:2015jjp, Bellm:2017bvx, Bellm:2019zci}, detector simulation/analysis through the \texttt{HwSim} package~\cite{hwsim} and signal and background separation, we refer the reader to Appendix F of ref.~\cite{Papaefstathiou:2020iag}. 

\subsection{\boldmath $pp \rightarrow ZZ \rightarrow (\ell^+ \ell^-) (\ell'^+ \ell'^-)$}
In this analysis, events are considered if they contain four leptons with transverse momenta satisfying, from hardest to softest, at least: $p_T(\ell_{1,2,3,4}) > 50, 50, 30, 20$~GeV. Further, events are only accepted if they contain two pairs of oppositely-charged same-flavour leptons and these are combined to form the $Z$ boson candidates, with the constraint $m(\ell \ell) \in [12, 120]$~GeV. If an event does not contain at least two $Z$ boson candidates, it is rejected. In the case of four same-flavour leptons, if there exist two viable lepton combinations, the combination $(\ell_i \ell_j)(\ell_k \ell_l)$ with the lowest value of $\chi^2 = (m(\ell_i \ell_j)-m_Z)^2 + (m(\ell_k \ell_l) - m_Z)^2$ is chosen, forming the candidates $Z_1$ and $Z_2$. We require that the combined invariant mass of the four leptons satisfies $m(\ell_i \ell_j\ell_k \ell_l) > 180$~GeV. 

We construct a set observables consists that we then feed into a boosted decision tree (BDT) via the \texttt{ROOT} \texttt{TMVA} package~\cite{Speckmayer:2010zz}. This set consists of the lepton transverse momenta, $p_T(\ell_{1,2,3,4})$, the combined lepton invariant mass $m(\ell_i \ell_j\ell_k \ell_l)$, the transverse momentum of the two $Z$ boson candidates, $p_T(Z_1)$, $p_T(Z_2)$, their invariant masses, $m(Z_1)$, $m(Z_2)$, their distance $\Delta R(Z_1, Z_2)$, the distance between the leptons that form the two candidates, $\Delta R(\ell_i, \ell_j)$ and $\Delta R(\ell_k \ell_l)$, the invariant mass of the combined $Z$ boson candidates $m(Z_1 Z_2)$ and their combined transverse momentum, $p_T(Z_1 Z_2)$. 

We consider only the dominant backgrounds, originating from non-resonant and resonant SM four lepton production, matched at NLO via the \texttt{MC@NLO} method~\cite{Frixione:2002ik}. In addition, we consider the LO gluon-fusion component of four lepton production that originates from the resonant loop-induced production of two $Z$ bosons, i.e.\ $gg \rightarrow ZZ$, deemed to be important at higher proton-proton centre-of-mass energies~\cite{Harlander:2018yns}. 

\subsection{\boldmath $pp \rightarrow h_1 h_1 \rightarrow (b\bar{b}) (\gamma \gamma)$}
 We require all jets (including $b$-tagged) to have transverse momentum $p_T > 30$~GeV and to lie within $|\eta| < 3.0$. The $b$-jet tagging probability was set to 0.75, uniform over the transverse momentum. The jet to photon mis-identification probability was set to $0.01 \times \exp{(p_T/30~\mathrm{GeV})}$, where $p_T$ is the jet transverse momentum~\cite{ATL-PHYS-PUB-2013-009}. We require that the invariant mass of the two $b$-jets lies in $m_{bb} \in [100, 150]$~GeV and that the invariant mass of the di-photon system within $m_{\gamma\gamma} \in [115, 135]$~GeV. 

The final set of observables constructed for the BDT consists of:  the invariant mass of the two $b$-jets, $m_{bb}$,  the invariant mass of the di-photon system, $m_{\gamma\gamma}$, the invariant mass of the combined system of the two $b$-jets and the photons, $m_{bb\gamma\gamma}$, the distance between the $b$-jets, $\Delta R (b,b)$, the distance between the photons, $\Delta R (\gamma \gamma)$, the distance between the two $b$-jet system and the di-photon system, $\Delta R (b b, \gamma \gamma)$,  the transverse momentum of each $b$-jet, $p_T(b_1,2)$, the transverse momentum of each photon $p_T(\gamma_1)$, $p_T(\gamma_2)$, the transverse momentum of the two $b$-jet system, $p_T(bb)$, the transverse momentum of the di-photon system $p_T(\gamma\gamma)$, the transverse momentum of the combined $b$-jet and photon systems, $p_T(bb\gamma\gamma)$ and the distances between any photon and any $b$-jet, $\Delta R (b_i, \gamma_j)$ with $i, j=1,2$. 

As backgrounds we consider $\gamma\gamma$+jets, $\gamma$+jets, by producing, respectively, $\gamma\gamma j$ and $\gamma j j$ via MC@NLO, $t\bar{t} \gamma \gamma$ via MC@NLO, $b\bar{b}\gamma\gamma$ and $bj\gamma\gamma$ at LO. We also consider backgrounds originating from single Higgs boson production: $b\bar{b} h_1$, $Z h_1$, $t\bar{t} h_1$, where we assume that the branching ratios possess their SM values. As an approximation, we also consider the non-resonant part of $h_1 h_1$ as a background, assuming that the self-coupling maintains a value close to the SM value.

\subsection{\boldmath $pp \rightarrow ZZ \rightarrow (\ell^+ \ell^-) (\nu \nu)$}

The event selection for the  $pp \rightarrow ZZ \rightarrow (\ell^+ \ell^-) (\nu \nu)$ final state consists of combining di-lepton $Z$ boson candidates with a relatively large missing transverse momentum ($\slashed{p}_T$). We require two oppositely-charged leptons of the same flavour, each with $p_T(\ell) > 50$~GeV. We further require their combined invariant mass within 30~GeV of the $Z$ boson mass and di-lepton transverse momentum, $p_T(\ell\ell) > 55$~GeV. In addition require $p_T^{\mathrm{miss}} > 125$~GeV. We veto events if $\Delta \phi(\vec{\slashed{p}}_T,~\mathrm{any~jet~with~}p_T > 30\mathrm{~GeV}) < 0.5$, where $\Delta \phi$ is the difference in angle between the $\vec{\slashed{p}}_T$ and any jet on the plane perpendicular to the beam axis. We also require the $Z$ boson candidate to satisfy $\Delta \phi (Z, \vec{\slashed{p}}_T) > 0.5$. We construct the transverse mass as:
\begin{equation}\label{eq:mt2l2v}
m_T^2 = \left( \sqrt{ p_T(\ell\ell)^2 + m(\ell \ell)^2} + \sqrt{ \slashed{p}_T^2 + m_Z^2} \right)^2 - (\vec{p}_T(\ell \ell) + \vec{\slashed{p}}_T)^2 \;,
\end{equation}
where $m(\ell \ell)$ is the invariant mass of the di-lepton system. This is employed in the present article to obtain an estimate of the mass of the $h_2$, since $m_T \leq m_2$.

The final set of observables that are used in the discrimination of signal versus background consists of: the transverse momenta of the leptons that form the $Z$ boson candidate, $p_T(\ell_1)$, $p_T(\ell_2)$, the corresponding, di-lepton invariant mass, $m(\ell\ell)$, and transverse momentum, $p_T(\ell \ell)$, their pseudo-rapidity distance $\Delta \eta = | \eta (\ell_1) - \eta (\ell_2) |$ and their distance $\Delta R = \sqrt{ \Delta \eta^2 + \Delta \phi^2 }$, the transverse mass $m_T$ as defined above and the magnitude of the missing transverse momentum, $\slashed{p}_T$. 

As backgrounds, we consider those that can yield the $2\ell$ final state with an associated missing transverse momentum, originating from the on-shell production of $ZZ$, $WZ$, $ZVV$ where $V=W,Z$, $t\bar{t}$ and $WW$ production, all matched via the \texttt{MC@NLO} method to the parton shower. We do not consider the mis-identification of jets or photons as leptons, and we do not include $\tau$ leptons in either signal or backgrounds.

\section{Calculating the uncertainty on the cross section given the significance}\label{app:xsecuncert}
For the sake of completeness, we discuss here the origin of eq.~\ref{eq:xsecuncert},
\begin{equation}
\Delta \sigma(xx) = \frac{\sigma(xx)}{\Sigma(xx)}\;,
\end{equation}
used to estimate the uncertainty in the measurement of the signal cross section at a given statistical significance. We consider the calculation of the number of signal events $S$ given an observed number of events $N$ and expected background events $B$. We assume that the $N$ events follow a gaussian distribution.\footnote{We note that in ref.~\cite{Papaefstathiou:2020iag}, we imposed that the number of signal events in \textit{any} channel should be greater than 100, therefore justifying the gaussian approximation.} To estimate the number of signal events in a given sample, one has to subtract the expected number of background events from $N$, i.e.\ $S=N-B$. The statistical error on this estimate is simply $\Delta S = \sqrt{N}$. But $N=S+B$, therefore $\Delta S = \sqrt{S+B}$. Now, $\Delta S / S = \sqrt{S+B}/S$. The expression $\Sigma = S/\sqrt{S+B}$ is an estimate of the gaussian statistical significance, and therefore $\Delta S / S = 1/\Sigma$. Since an estimate of the cross section is obtained by rescaling $S$ by the collider integrated luminosity, $\Delta \sigma / \sigma = 1/\Sigma$ and we obtain eq.~\ref{eq:xsecuncert}.

\bibliographystyle{JHEP}
\bibliography{references.bib}

\end{document}